\documentclass{article}

\usepackage{arxiv}

\usepackage[utf8]{inputenc} 
\usepackage[T1]{fontenc}    
\usepackage{hyperref}       
\usepackage{url}            
\usepackage{booktabs}       
\usepackage{amsfonts}       
\usepackage{nicefrac}       
\usepackage{microtype}      
\usepackage{lipsum}         
\usepackage{graphicx}
\usepackage{natbib}
\usepackage{doi}

\usepackage{mathtools}
\usepackage{amsmath}
\usepackage{bm}
\usepackage{multirow}

\usepackage{cleveref}       
\title{Weakly nonlinear analysis of the onset of convection in rotating spherical shells}

\usepackage{authblk}

\author[1]{%
	\href{https://orcid.org/0000-0003-0994-2013}{Calum S. Skene}\thanks{Corresponding author: \texttt{c.s.skene@leeds.ac.uk}}%
}
\author[1]{%
	\href{https://orcid.org/0000-0003-0205-7716}{Steven M. Tobias}%
}
\affil[1]{Department of Applied Mathematics, University of Leeds, Leeds LS2 9JT, UK}

\renewcommand{\shorttitle}{Weakly nonlinear analysis of rotating convection}

\hypersetup{
pdftitle={Weakly nonlinear analysis of the onset of convection in rotating spherical shells},
pdfsubject={},
pdfauthor={Calum S. Skene, Steven M. Tobias},
pdfkeywords={weakly nonlinear analysis, rapidly rotating convection, planetary dynamos, adjoints},
}

\begin{document}
\maketitle

\begin{abstract}
A weakly nonlinear study is numerically conducted to determine the behaviour near the onset of convection in rotating spherical shells. The mathematical and numerical procedure is described in generality, with the results presented for an Earth-like radius ratio. Through the weakly nonlinear analysis a Stuart--Landau equation is obtained for the amplitude of the convective instability, valid in the vicinity of its onset. Using this amplitude equation we derive a reduced order model for the saturation of the instability via nonlinear effects and can completely describe the resultant limit cycle without the need to solve initial value problems. In particular the weakly nonlinear analysis is able to determine whether convection onsets as a supercritical or subcritical Hopf bifurcation through solving only linear 2D problems, specifically one eigenvalue and two linear boundary value problems. Using this, we efficiently determine that convection can onset subcritically in a spherical shell for a range of Prandtl numbers if the shell is heated internally, confirming previous predictions. Furthermore, by examining the weakly nonlinear coefficients we show that it is the strong zonal flow created through Reynolds and thermal stresses that determines whether convection is supercritical or subcritical.
\end{abstract}


\section{Introduction}

Many planets in our solar system such as the Earth, Mercury, Jupiter, and Saturn, as well as Jupiter's moon Ganymede, exhibit magnetic fields (see the review of \citet{Jones_2011} for example). These planetary bodies all consist of conducting fluid cores; with liquid metal in Mercury, Earth, and Ganymede, and metallic hydrogen in the gas giants Jupiter and Saturn. Due to this, dynamo action - by which a magnetic field arises and is sustained by the motion of a conducting fluid - is widely believed to be responsible for maintaining these field. Without such a mechanism sustaining these fields, any magnetic field present when they formed would have long since diffused. In all these planetary dynamo examples, convection is believed to be the energy source providing the motion of the conducting fluid. Therefore, it is of importance to investigate convection in these systems in order to understand the mechanisms behind planetary dynamos.

Owing to the critical role that convection plays in the planetary dynamo process, a lot of work has been conducted in order to understand how convection arises in rapidly rotating spherical systems \citep{Proctor_1994,Busse_2002}. \citet{Roberts1965} introduced the asymptotic theory for the onset of convection in spheres in the rapidly rotating (low Ekman number) limit, relevant to planetary bodies. Subsequently, asymptotic solutions were proposed by \citet{Roberts1968} and \citet{Busse_1970}, giving a local stability criterion on the Rayleigh number for the onset of instability. However, these solutions fail to capture the global instability through the process of phase-mixing, whereby local disturbances that initially grow exponentially are moved away from their region of instability and thereafter decay away \citep{soward_1977}. The true asymptotic theory for the onset of convection in spheres was given by \citet{Jones_2000}, where the complication of phase mixing is addressed through the method outlined by \citet{Yano_1992}. This gives the global critical Rayleigh number for the onset of convection, and differs by an $\mathcal{O}(1)$ amount from the local critical value of Roberts and Busse. Following on from the work of \citet{Jones_2000}, the asymptotic theory of the onset of convection in rapidly rotating spherical shells was given by \citet{dormy_2004}. 

The nature of the onset of convection {for rapid rotation} revealed by these studies is as follows. Convection first onsets as a thermal Rossby wave, which due to the Taylor--Proudman constraint takes the form of a number of columns aligned with the rotation axis with azimuthal wavenumber $m$. As the Ekman number is lowered, larger Rayleigh numbers and azimuthal wavenumbers are {required for} the onset of convection. In the case of a {full} sphere, internal heating is used in order to provide a heat difference between the centre of the sphere and the sphere's boundary, and convection onsets at a critical cylindrical radius. In contrast, in the case of a shell two different forms of heating {have been} considered by  \citet{dormy_2004}. Differential heating directly imposes a {temperature} difference between the inner and outer boundaries of the shell, whereas internal heating {also includes uniformly distributed internal heat sources}. \citet{dormy_2004} show that an internally heated shell behaves similarly to a full sphere provided the radius ratio is small enough. However, for differential heating, convection always onsets at a cylindrical radius adjacent to the inner shell boundary. In addition to the numerical solutions given for an Earth-like radius ratio in \citet{dormy_2004}, the more recent study by \citet{Barik2023} considers the onset of convection in a shell for a variety of Ekman numbers and radius ratios. In this manner, a database of critical Rayleigh numbers, critical wavenumbers, and angular velocities of the thermal Rossby waves present at the onset of convection was made available.

In {both} the {full} sphere and shell, convection onsets as a Hopf bifurcation, whereby a complex conjugate pair of eigenvalues transition from being stable to unstable at the critical Rayleigh number. However, these studies do not reveal the nature of this bifurcation. In a supercritical Hopf bifurcation nonlinearities saturate the instability to form a stable periodic limit cycle {immediately} after onset. On the other hand, in a subcritical Hopf bifurcation nonlinearities promote the growth of the instability, leading to a limit cycle solution below the onset of instability (which can be reached with a finite amplitude perturbation). In numerical simulations convection is most often found to onset supercritically (see the benchmark of \citet{Christensen2001} for example). However, the weakly nonlinear analyses of \citet{soward_1977} and \citet{plaut_lebranchu_simitev_busse_2008} suggest that this onset could change to being subcritical for small enough Ekman number for the case of internal heating. In the full sphere this change to a subcritical bifurcation has been confirmed by \citet{Guervilly_2016} and \citet{Kaplan2017} for low Prandtl numbers. In these studies the fully nonlinear equations are solved near onset in order to determine whether the instability is supercritical or subcritical. Due to the the low Ekman numbers required to find a subcritical bifurcation (at least $\textit{Ek}\approx 10^{-6}$), these studies have high numerical requirements necessitating the use of a quasi-geostrophic model model in \citep{Guervilly_2016} and hyperviscosity in \citep{Kaplan2017}.

A {more efficient} method for  determining whether a bifurcation is supercritical or subcritical is weakly nonlinear analysis \citep{Malkus_Veronis_1958, Stuart_1960,Watson_1960}. In a weakly nonlinear analysis the critical parameter at which an instability arises is first found, and then perturbed by an $\mathcal{O}(\epsilon^2)$ amount. At order $\mathcal{O}(\epsilon)$ the system resembles the original system at onset, however at $\mathcal{O}(\epsilon^2)$ select, weak, nonlinearities enter the picture. The key idea is that by allowing the instability to evolve on an $\mathcal{O}(\epsilon^2)$ timescale, a Stuart--Landau equation \citep{Landau_1994} for the amplitude of the instability can be determined at $\mathcal{O}(\epsilon^3)$. As this equation is the normal form for a Hopf bifurcation, it is then immediately apparent from the coefficients what the nature of the instability is. As all the equations in the weakly nonlinear analysis remain linear, it has become an important and powerful mathematical technique for extending linear stability analyses. 

In spherical convection problems weakly nonlinear analysis has been used to show the formation of differential rotation in the convection zone of stars \citep{Busse_1970b,Busse_1973}. This differential rotation, or strong zonal flow, is a key feature of rotating convection confirmed by numerical simulations \citep{Gilman1977,Gilman_1978,Tilgner_Busse_1997}. In non-rotating convection \citet{mannix_mestel_2019} use weakly nonlinear analysis to study a degenerate point where two modes become unstable simultaneously in spherical convection. Perhaps of most relevance to our current work is the weakly nonlinear analysis presented by \citet{plaut_lebranchu_simitev_busse_2008} who perform weakly nonlinear analysis on a 2D quasi-geostrophic model of rotating convection in a spherical shell. In this study the weakly nonlinear analysis showed the formation of a strong zonal-flow due to Reynolds stress terms, and aligned with predictions \citep{soward_1977} that convection in a spherical shell becomes subcritical at low enough Ekman numbers for internal heating.

In this study we seek to extend linear stability analyses for determining the onset of convection in a spherical shell by performing weakly nonlinear analysis numerically with no simplifications of the governing equations. This methodology follows recent studies {that} have derived the weakly nonlinear equations analytically {for complex non self-adjoint systems}, and solved them numerically. {This has allowed} otherwise intractable problems, such as flow over an open cavity and flow past a cylinder \citep{sipp_2012} and a swirling jet \citep{Skene2020}, to be tackled. In this manner, a reduced order model can be found for the onset of convection accurate for Rayleigh numbers close to the critical value. Through this reduced description, the limit cycle solution produced via instability can be determined without solving any initial value problems. By examining the structure of this limit cycle, its amplitude can be determined as a function of Rayleigh number, and nonlinear effects such as a change in the angular rotation rate can be found. Crucially, the nature of the Hopf bifurcation can be determined, allowing for an efficient search to be conducted for the parameters at which convection onsets subcritically in a spherical shell. The rest of the paper is organised as follows; \S \ref{sec:maths} presents the mathematical formulation behind the weakly nonlinear analysis, \S \ref{sec:numerics} discusses the numerical procedure, the results are contained in \S \ref{sec:results}, and conclusions are offered in \S \ref{sec:conclusions}.

\section{Mathematical formulation}\label{sec:maths}
We consider a spherical shell with inner and outer radius $r_i$ and $r_o$, respectively. The non-dimensional Navier--Stokes equations under the Boussinesq approximation are
\begin{gather}
\frac{\partial \bm{u}}{\partial t}+\bm{u}\bm{\cdot}\nabla\bm{u}= -\nabla p -2\bm{e}_z\times\bm{u} + \tilde{\textit{Ra}}\textit{Ek} \frac{r}{r_0}T\hat{\bm{e}}_r +\textit{Ek} \nabla^2\bm{u},\\
\nabla \bm{\cdot} \bm{u}=0,\\
\frac{\partial T}{\partial t}+\bm{u}\bm{\cdot}\nabla T=\frac{\textit{Ek}}{\textit{Pr}}\left( \nabla^2 T + S \right),
\end{gather}
for the fluid velocity $\bm{u}$, pressure $p$, and temperature $T$. These equations have been non-dimensionalised as follows; length using the shell thickness $d=r_0-r_i$, temperature with the temperature difference across the shell $\Delta T$, time with the rotational time $\Omega$, and velocities with $d\Omega$. This introduces the non-dimensional parameters of the \textit{modified} Rayleigh number $\tilde{\textit{Ra}}=\textit{Ra}\textit{Ek}/\textit{Pr}$, Ekman number $\textit{Ek}=\nu/(\Omega d^2)$, and Prandtl number $\textit{Pr}=\nu/\kappa$. Note that the modified Rayleigh number is used rather than the traditional Rayleigh number $\textit{Ra}=\tilde{\textit{Ra}}\textit{Ek}/\textit{Pr}$ to account for the stabilising effect of lowering the Ekman number. {Following \citet{dormy_2004}}, the temperature equation has a source term $S$ which takes the form
\begin{equation}
    S = 
    \begin{dcases}
    0 & \textrm{Differential heating}\\
    6\frac{1-\beta}{1+\beta} & \textrm{Internal heating} \\
    \end{dcases},
\end{equation}
with $\beta = r_i/r_o$ the radius ratio which is fixed at the Earth-like value of $\beta=0.35$ for the rest of the study. Boundary conditions are no-slip for the fluid velocity $\bm{u}(r=r_i)=\bm{u}(r=r_o)=0$, and fixed temperature on the shell walls $T(r=r_i)=1$, and $T(r=r_o)=0$.

\begin{figure}
	\centering
	\includegraphics[width=0.48\textwidth]{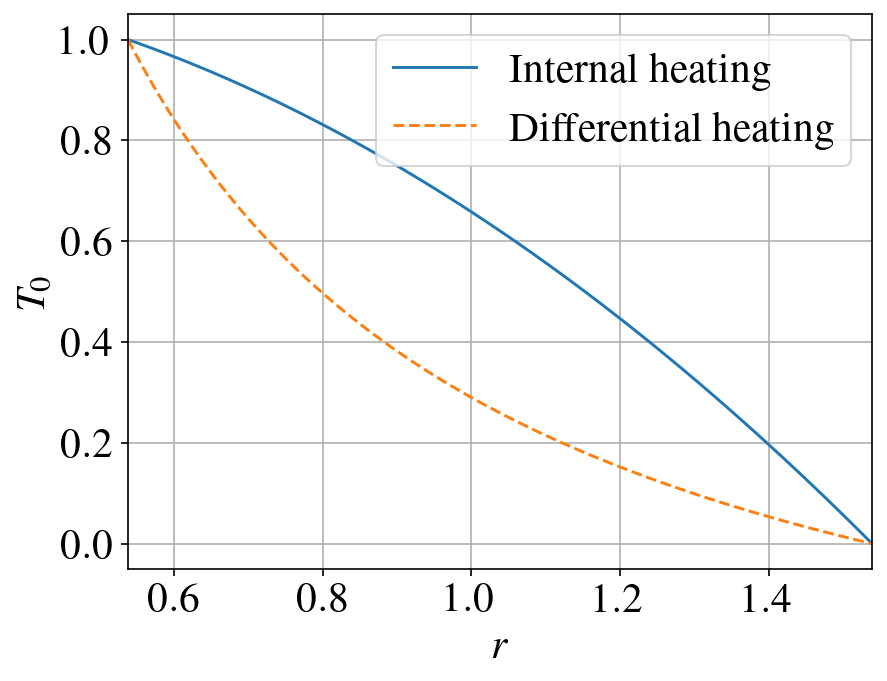}%
    \includegraphics[width=0.48\textwidth]{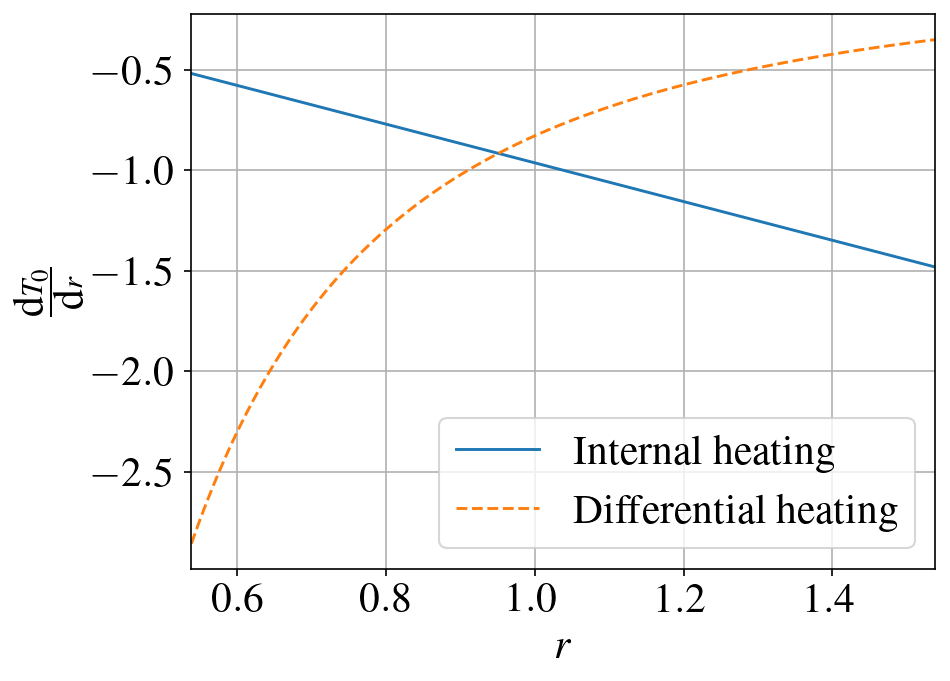}
	\caption{Temperature (left) and temperature gradient (right) profiles for internal heating and differential heating.}
    \label{fig:temp_profiles}
\end{figure}

We begin our analysis by seeking a solution in the form $\bm{q}=\bm{q}_0(r) + \epsilon A\bm{q}_A(\theta,r)\exp(i \omega_c t +im_c\phi) +\textrm{c.c.}$, at $\tilde{\textit{Ra}}=\tilde{\textit{Ra}}_c$ and with $\epsilon\ll 0$. As $\epsilon$ is small, substituting this form of the solution into the equations separates into two separate problems. The first is that $\bm{q}_0$ is a steady solution, which has solution $\bm{q}_0(r)=(\bm{u}_0,p_0,T_0)^T=(\bm{0},p_0,T_0)^T$, {called the conducting state}, where
\begin{equation}
    \frac{\textrm{d} T_0}{\textrm{d} r}=
    \begin{dcases}
    -\frac{r_ir_o}{(r_o-r_i)^2}r^{-2} & \textrm{Differential heating} \\
    -2\frac{1-\beta}{1+\beta}r & \textrm{Internal heating} \\
    \end{dcases},
\end{equation}
and $p_0$ balances the buoyancy term in the momentum equation. Temperature profiles of the steady base {state} are shown in figure \ref{fig:temp_profiles}. In both cases the temperature monotonically decreases from the inner radius to the outer radius of the shell. The key difference is the gradient of this decrease. For internal heating the temperature gradient is linear, whereas for the case of differential heating the temperature gradient has a $r^{-2}$ dependence, causing it to be much more negative close to the inner shell boundary. 

The second problem is the eigenvalue problem $i\omega_c\mathcal{M}\bm{q}_A=\mathcal{L}_{m_c}\bm{q}_A$, where
\begin{equation}
    \mathcal{L}_m\bm{q}=
    \begin{pmatrix}
-\nabla_m p -2\bm{e}_z\times\bm{u} + \tilde{\textit{Ra}}_c\textit{Ek}\frac{r}{r_0}T\hat{e}_r +\textit{Ek} \nabla_m^2\bm{u} \\
\nabla_m \bm{\cdot} \bm{u} \\
-\bm{u}_A\bm{\cdot}\nabla_m T_0+\frac{\textit{Ek}}{\textit{Pr}} \nabla_m^2 T_A\\
    \end{pmatrix},
    \label{equ:eigenvalue}
\end{equation}
and
\begin{equation}
    \mathcal{M}=
    \begin{pmatrix}
    \bm{1} & 0 & 0 \\
    0 & 0 & 0 \\
    0 & 0 & 1
    \end{pmatrix}.
\end{equation}
Note that we have introduced the operator $\nabla_m$ which is the same as $\nabla$, except that $\phi$ derivatives are replaced with multiplications by $im$. This eigenvalue problem, {for loss of stability in a Hopf bifurcation}, requires that the eigenvalue is purely imaginary, and is achieved when $\tilde{\textit{Ra}}$ is the critical Rayleigh number $\tilde{\textit{Ra}}_c$. We now consider Rayleigh numbers close, but not equal, to this critical value. Specifically we will consider solutions at the Rayleigh number $\tilde{\textit{Ra}}=\tilde{\textit{Ra}_c}+\epsilon^2\tilde{\textit{Ra}_c}$. At this Rayleigh number we do not change the analysis already presented as we are only introducing perturbations at $\mathcal{O}(\epsilon^2)$. This definition gives the small parameter $\epsilon$ as the relative distance from the critical modified Rayleigh number, i.e.
$
    \epsilon^2 = (\tilde{\textit{Ra}}-\tilde{\textit{Ra}}_c)/\tilde{\textit{Ra}}_c.
$
However, we are now above the critical Rayleigh number and so should expect the perturbation to grow. This can be quantified by finding an equation for $A(t)$ which is the goal of the weakly nonlinear analysis we now present. 

In order to find an amplitude equation we must go beyond $\mathcal{O}(\epsilon)$. To do this we can seek a solution in the form 
\begin{gather}
\bm{q}=\bm{q}_0(r) + \epsilon \left[A\bm{q}_A(\theta,r)\exp(i \omega_c t +im_c\phi) + \textrm{c.c.} \right] + \epsilon^2( A\bar{A}\bm{q}_{A\bar{A}}(\theta,r)+ \label{equ:second_order}\\A\bar{A}\bm{q}_{\textit{Ra}}(\theta,r)
+ [AA\bm{q}_{AA}(\theta,r)\exp(2i \omega_c t +2im_c\phi) + \textrm{c.c.}]) + \mathcal{O}(\epsilon^3)\nonumber
\end{gather}
Substituting this into the governing equations gives the same equations we have already considered at $\mathcal{O}(0)$ and $\mathcal{O}(\epsilon)$. At $\mathcal{O}(\epsilon^2)$ we obtain
\begin{gather}
(2i\omega_c\mathcal{M} -\mathcal{L}_{2m_c})\bm{q}_{AA} =
(
-\bm{u}_A\bm{\cdot}\nabla_{m_c}\bm{u}_A ,
0,
-\bm{u}_A\bm{\cdot}\nabla_{m_c}T_A
)^T,
\label{equ:qAA}
\\
-\mathcal{L}_0\bm{q}_{A\bar{A}} =
(
-\bm{u}_A\bm{\cdot}\nabla_{-m_c}\bm{u}_{\bar{A}} - \bm{u}_{\bar{A}}\bm{\cdot}\nabla_{m_c}\bm{u}_A ,
0,\label{equ:qAAb}
-\bm{u}_A\bm{\cdot}\nabla_{-m_c} T_{\bar{A}} - \bm{u}_{\bar{A}}\bm{\cdot}\nabla_{m_c} T_A )^T,
\\
-\mathcal{L}_0\bm{q}_{\textit{Ra}} =
\left( \tilde{\textit{Ra}}_c\textit{Ek} \frac{r}{r_0} T_0 \hat{\bm{e}}_r , 0,0\right)^T.
\label{equ:qRa}
\end{gather}
Solving these determines the second order corrections $\bm{q}_{AA}$ and $\bm{q}_{A\bar{A}}$, stemming from nonlinear interactions of the neutral eigenvector, as well as base-flow modification $\bm{q}_{\textit{Ra}}$ that arises from increasing the Rayleigh number. Notably they all take the form of resolvent problems. In other words, the harmonics and base-flow modifications are all found as the long-term driven response to forcings arising from nonlinear interactions.

To determine an equation for the amplitude $A$ we first make the assumption that as we are only just supercritical that the growth of the eigenmode will be slow. This is quantified by writing that the amplitude $A$ depends on the slow time scale $T=\epsilon^2 t$, i.e. $A:=A(T)$. At third order we then obtain the following equation at $m=m_c$ and $\omega=i\omega_c$
\begin{equation}
(i\mathcal{M}\omega_c-\mathcal{L}_{m_c})\bm{q}_3=-\mathcal{M}\bm{q}_A\frac{d A}{d T} + \left[\boldsymbol{\chi}A - \boldsymbol{\gamma}A|A|^2\right],
\end{equation}
where
\begin{gather}
\boldsymbol{\chi}=
\begin{pmatrix}
-\bm{u}_{\textit{Ra}}\bm{\cdot}\nabla_{m_c}\bm{u}_{A}-\bm{u}_{A}\bm{\cdot}\nabla_{0}\bm{u}_{\textit{Ra}} + \tilde{\textit{Ra}}_c\textit{Ek}\frac{r}{r_0}T_A\hat{\bm{e}}_r\\
0\\
-\bm{u}_{\textit{Ra}}\bm{\cdot}\nabla_{m_c}T_{A}-u_{A}\bm{\cdot}\nabla_{0}T_{\textit{Ra}} \\
\end{pmatrix},
\label{equ:chi}
\\
\boldsymbol{\gamma}=
\begin{pmatrix}
\bm{u}_{A\bar{A}}\bm{\cdot}\nabla_{m_c}\bm{u}_{A}+\bm{u}_{A}\bm{\cdot}\nabla_{0}\bm{u}_{A\bar{A}}
+\bm{u}_{AA}\bm{\cdot}\nabla_{-m_c}\bm{u}_{\bar{A}}+\bm{u}_{\bar{A}}\bm{\cdot}\nabla_{2m_c}\bm{u}_{AA}\\
0\\
\bm{u}_{A\bar{A}}\bm{\cdot}\nabla_{m_c}T_{A}+\bm{u}_{A}\bm{\cdot}\nabla_{0}T_{A\bar{A}}
+\bm{u}_{AA}\bm{\cdot}\nabla_{-m_c}T_{\bar{A}}+\bm{u}_{\bar{A}}\bm{\cdot}\nabla_{2m_c}T_{AA}\\
\end{pmatrix},
\label{equ:gamma}
\end{gather}
and $\bm{q}_3$ is a third order term. Ensuring that this equation is solvable, and hence our weakly nonlinear expansion is valid, will yield the amplitude equation. In order to do this we first introduce the inner product $\langle \bm{\cdot},\bm{\cdot}\rangle$ and define adjoint operators $\mathcal{L}_m^\dagger$ and $\mathcal{M}^\dagger$ as the operators that satisfy $\langle \bm{a},\mathcal{L}_m\bm{b}\rangle=\langle \mathcal{L}_m^\dagger\bm{a},\bm{b}\rangle$ and $\langle \bm{a},\mathcal{M}\bm{b}\rangle=\langle \mathcal{M}^\dagger\bm{a},\bm{b}\rangle$, for all $\bm{a}$ and $\bm{b}$, respectively. The Fredholm alternative then tells us that this equation will be solvable when
\begin{equation}
\langle\bm{q}_A^\dagger,\mathcal{M}\bm{q}_A\rangle\frac{d A}{d T} = \chi A - \gamma A|A|^2,
\end{equation}
with $\chi=\langle\bm{q}_A^\dagger,\boldsymbol{\chi}\rangle$ and $\gamma=\langle\bm{q}_A^\dagger,\boldsymbol{\gamma}\rangle$. The vector $\bm{q}_A^\dagger$ solves the adjoint eigenvector problem
\begin{equation}  \mathcal{L}_m^\dagger\bm{q}_A^\dagger=-i\omega_c\mathcal{M}^\dagger \bm{q}_A^\dagger.
\end{equation}
We note that there are many choices of inner product $\langle \bm{\cdot},\bm{\cdot}\rangle$, with different choices leading to different adjoint eigenvalue problems. However, all choices will lead to the same values of $\chi$ and $\gamma$, therefore a numerically convenient inner product can be chosen. The choice for our study will therefore be discussed in the next section on the numerical implementation. For convenience in what follows we will write $\chi=\chi_r+i\chi_i$ and $\gamma=\gamma_r+i\gamma_i$, where $\chi_r,\chi_i,\gamma_r,\gamma_i\in\mathbb{R}$.

By normalising the adjoint eigenvector such that $\langle\bm{q}_A^\dagger,\mathcal{M}\bm{q}_A\rangle=1$ we obtain the simplified amplitude equation 
\begin{equation}
\frac{d A}{d t} = \epsilon^2(\chi A - \gamma A|A|^2).
\end{equation}
This is a Stuart--Landau equation governing the growth of the eigenmode. By also normalising $\bm{q}_A$ to have unit energy, we have that the energy of the instability will grow as $E(t)=E_0\epsilon^2|A(t)|$.  For small $A$ the solution behaves like $A\approx\exp(\epsilon^2\chi t)$, showing that if $\textrm{Re}(\chi)>0$ we will get exponential growth (we will assume $\textrm{Re}(\chi)>0$ for the rest of the paper). When $A$ grows larger the cubic nonlinearities become important.

To explore the role of the cubic nonlinearity we can write the complex amplitude in the form $A=\eta\exp(i\psi)$, where $\eta$ is the amplitude and $\psi$ is the phase. This leads to the polar form of the amplitude equation
\begin{gather}
\frac{\textrm{d}\eta}{\textrm{d}t} = \epsilon^2 \left(\chi_r \eta - \gamma_r \eta^3\right),\\
\frac{\textrm{d}\psi}{\textrm{d}t} = \epsilon^2 \left(\chi_i  - \gamma_i \eta^2\right).
\end{gather}
From this form we see that there are potentially two steady solutions to the $\eta$-equation. The first is $\eta=0$, corresponding to our fixed point. The second is when $\gamma_r>0$, which has amplitude $\eta_\textrm{LC}=\sqrt{\chi_r/\gamma_r}$, and phase $\psi_\textrm{LC} = \psi_0+t\epsilon^2(\chi_i-\gamma_i\chi_r/\gamma_r)=\psi_0+t\epsilon^2 \psi_\epsilon$. In this manner, we can see that if $\gamma_r>0$ the instability is a \textit{supercritical} Hopf bifurcation, i.e. the nonlinearities saturate the growing instability to form a stable limit cycle with the amplitude and phase described by the previous expressions. On the other hand, if $\gamma_r<0$ then nonlinearities promote the instability and no stable limit cycle is reached. This is the case of a \textit{subcritical} Hopf bifurcations. Hence, we can use the weakly nonlinear theory to describe the limit cycle near criticality without timestepping either the Stuart--Landau amplitude equation or full governing equations, {just by determining the sign of $\gamma_r$. This comes at the expense of finding $\gamma_r$ which is not an inconsiderable problem numerically.}

Using the limiting expressions for $A$ we can substitute them into our solution to see that as $t\rightarrow\infty$
\begin{equation}
    \bm{q}\rightarrow  \bm{q}_0+\epsilon A_{\textrm{LC}}\bm{q}_A\exp((i\omega_c+\epsilon^2\psi_\epsilon) t+i\phi m)+\textrm{c.c.}+\mathcal{O}(\epsilon^2).
    \label{equ:limitCycle}
\end{equation}
This is a second order accurate expression for the limit cycle. A third order accurate expression could also have been obtained by retaining terms up to second order, introducing the second order harmonics and base-flow modifications into the expression. Equation (\ref{equ:limitCycle}) shows that the limit cycle has its phase modified from that of the eigenvalue to $\psi_\textrm{LC} = i\omega_c+\epsilon \psi_\epsilon$.
We can also use equation (\ref{equ:limitCycle}) to calculate the energy of the limit cycle as
\begin{equation}
E = \frac{1}{2}\iiint \bm{u}\bm{\cdot} \bm{u}\;\textrm{d}V \approx 4\epsilon^2A_{\textrm{LC}}^2 + \mathcal{O}(\epsilon^4).
\label{equ:energy}
\end{equation}
As well as giving a direct link between $A_{\textrm{LC}}$ and $E$, this equation gives us a way to verify the weakly nonlinear setup. By performing a few nonlinear simulations of the full governing equations at different values of $\epsilon$ we can check that the saturated limit-cycle energy $E$ matches that predicted by equation (\ref{equ:energy}) with an error that decays as $\epsilon^4$.

Before discussing the numerics we first note that changing the Rayleigh number does not change the velocity or temperature components of the base-flow. Indeed, solving equation (\ref{equ:qRa}) will only give a pressure correction, with $\bm{u}_{\textit{Ra}}=\bm{0}$ and $T_{\textit{Ra}}=0$. This significantly simplifies the form of $\boldsymbol{\chi}$ given by equation (\ref{equ:chi}), so that $\chi$ only depends on the temperature component of the neutral eigenvector $T_A$. The result of this observation is that we only need to solve equations (\ref{equ:qAA}) and (\ref{equ:qAAb}) for $\bm{q}_{AA}$ and $\bm{q}_{A\bar{A}}$, respectively. In other words, after the critical Rayleigh number is determined only one additional adjoint eigenvector problem, as well as two linear boundary value problems, need to be solved to determine whether the instability is supercritical or subcritical. {We give details of this method in the next section.}

\section{Numerical implementation}\label{sec:numerics}

The open-source PDE solver {\tt Dedalus} \citep{Burns2020} is used to solve the eigenvalue problem, linear boundary value (resolvent) problems, as well as initial value problems needed for this study. Specifically, we use {\tt Dedalus v3} which has the capability of dealing with spherical domains \citep{Vasil2019,Lecoanet2019}.  {\tt Dedalus} solves equations using a spectral method. For example, for the eigenvalue problem (\ref{equ:eigenvalue}) {\tt Dedalus} will construct matrices $\bm{L}_m$ and $\bm{M}$ such that the discrete eigenvalue problem becomes
\begin{equation}
\bm{L}_m\hat{\bm{q}}=\lambda\bm{M}\hat{\bm{q}}.
\end{equation}
Note that $\hat{\bm{q}}$ is a vector of coefficients for the spectral representation of our state. A similar discretisation is performed for the linear boundary value problems. As {\tt Dedalus} explicitly constructs matrices for the eigenvalue problem, which can then be solved with \textit{scipy}'s {\tt ARPACK} wrapper, this motivates our choice of inner product. 

By defining our inner product between the spectral representation of two states $\bm{a}$ and $\bm{b}$ to be
\begin{equation}
\langle \hat{\bm{a}},\hat{\bm{b}}\rangle = \hat{\bm{a}}^H\bm{\hat{b}},
\end{equation}
we obtain the adjoint eigenvector equation
\begin{equation}
\bm{L}_m^H\hat{\bm{q}}^\dagger=\lambda^*\bm{M}^H\hat{\bm{q}}^\dagger.
\end{equation}
There are three main advantages to this formulation {(so-called discrete adjointing)}. Firstly, it is trivial to obtain the the adjoint eigenvector problem by Hermitian transpose of the matrices that {\tt Dedalus} constructs for the direct-eigenvector problem. In fact, {\tt Dedalus} already has this functionality built in. Secondly, the LU decomposition needed for finding eigenvalues using the shift-invert method can be shared between the direct and adjoint solves, significantly reducing the computational cost of additionally solving the adjoint eigenvalue problem. Thirdly, this discrete inner product implies that we are adjointing the discretised system, ensuring that we get the adjoint of our numerical model to machine precision (see the discussion in \citet{Mannix_2024}, for example). We note that by defining this inner product, we need to be consistent in its use. Specifically, it is this inner product that must be used when projecting the vector quantities to scalar quantities in forming the amplitude equation.

The first step in the weakly nonlinear analysis is to find the critical Rayleigh number $\tilde{\textit{Ra}}_c$ such that $\lambda=i\omega_c$, i.e. the eigenvalue is purely imaginary. {We note here a benefit of solving the adjoint equation even for the efficient solution of the linear problem}. The relation $\delta \lambda = \langle \hat{\bm{q}}^\dagger,\boldsymbol{\delta} \bm{L}_m\hat{\bm{q}}\rangle/\langle \hat{\bm{q}}^\dagger,\bm{M}\hat{\bm{q}}\rangle$ (see the review of \citet{Luchini_2014} for example), means that
\begin{equation}
\frac{\partial \lambda}{\partial \textit{Ra}} = \left\langle \hat{\bm{q}}^\dagger,\frac{\partial\bm{L}_m}{\partial \textit{Ra}}\hat{\bm{q}}\right\rangle,
\end{equation}
where we have simplified the expression using our chosen adjoint normalisation. In this manner, each time we calculate an eigenvalue at a given Rayleigh number, by also solving the adjoint eigenvector problem we have access to its derivative with respect to the Rayleigh number. Hence, by setting up an optimisation problem with cost functional $\mathcal{J}=\textrm{Real}(\lambda)^2$ we can use a gradient-based optimisation routine, here \textit{L-BFGS-B} from \textit{scipy}'s \textit{optimize} module, in order to converge our critical Rayleigh number.

\section{Results}\label{sec:results}
With the mathematical and numerical procedure described we now turn to the results, which we split into two sections. First, we consider the case of differential heating in \S \ref{sec:diff_heating}. This section illustrates the procedure outlined thus far for $\textit{Pr}=1$, and validates the weakly nonlinear formulation against solving the full nonlinear equations. Secondly, we consider internal heating in \S \ref{sec:internal_heating}. When considering internal heating we consider a range of Prandtl numbers $\textit{Pr}\in\{1,0.1,0.01\}$ and use the weakly nonlinear methodology to map out {efficiently} the parameter regime where convection onsets subcritically. {This would be exceptionally time consuming using timestepping of the full nonlinear equations.}

\subsection{Differential heating with $\textit{Pr}=1$}\label{sec:diff_heating}
Carrying out the weakly nonlinear procedure outlined in sections \ref{sec:maths} and \ref{sec:numerics}, for $\textit{Ek}\in \{10^{-3},10^{-4},10^{-5},10^{-6}\}$, results in the critical Rayleigh numbers and coefficients $\chi$ contained in table \ref{tab:eig_results}. The corresponding saturation terms $\gamma$ are displayed in table \ref{tab:coeff_diff}. {For each Ekman number,} we choose $m_c$ to be {the azimuthal wavenumber $m$ at which convection onsets for the smallest Rayleigh number}. We begin our results by analysing the behaviour at the large Ekman number $\textit{Ek}=10^{-3}$. At this Ekman number the lowest critical Rayleigh number is obtained for $m_c=4$. The optimisation procedure gives $\tilde{\textit{Ra}}_c\approx 55.9$, and an eigenvalue of $\lambda\approx -0.0231i$. The critical frequency shows that the first mode to become unstable, which takes the form of a thermal Rossby wave, travels prograde to the planet's rotation. As the growth rate is negligible, we can write $\lambda=i \omega_c$ with $\omega_c=-0.0231$. In addition to checking the growth rate is negligible, similarly to \citep{Barik2023} we also check the energy balance of the solution by defining the residual
\begin{equation}
\mathcal{R}=\frac{|D_\nu+D_e|}{\textrm{max}(\{|D_\nu|,|D_e|\})},
\end{equation}
where $D_\nu=\iiint_V \bm{u}_{A,r}\bm{\cdot}\nabla^2\bm{u}_{A,r}\;\textrm{d}V$ and $D_e=2\iiint_V \bm{e}:\bm{e}\;\textrm{d}V$. In defining the residual we have used $\bm{u}_{A,r}=\textrm{Real}(\bm{u}_A)$ and the rate of strain tensor $\bm{e}=1/2(\nabla\bm{u}_{A,r}+\nabla\bm{u}_{A,r}^T)$. By checking that $\mathcal{R}\approx 0$ we ensure that the resolution is sufficient to obtain converged results. Using the unstable eigenvector we can calculate the growth coefficient $\chi=0.039-0.0087i$. As the real part of this coefficient is positive, we can conclude that the instability will grow for $\textit{Ra}>\textit{Ra}_c$.

\begin{table}
\begin{center}
\def~{\hphantom{0}}
\begin{tabular}{ l c c c c c c c}\toprule
$\textit{Ek}$ & $\tilde{\textit{Ra}}_c$ & $m_c$ & $\omega_c$ & $\chi$ & $L_\textrm{max}$ & $N_\textrm{max}$ & $\mathcal{R}$\\[3pt]
\hline
$10^{-3}$ & $55.9$ & $4$ & $-0.0231$  & $3.88\times10^{-2}-8.66\times10^{-3}i$ & $43$ & $43$ & $1\times 10^{-12}$ \\
$10^{-4}$ & $75.2$ & $5$ & $-0.0112$  & $1.20\times10^{-2}-8.66\times10^{-3}i$ & $83$ & $83$ & $5 \times 10^{-14}$ \\
$10^{-5}$ & $106$ & $15$ & $-0.00712$ & $8.60\times10^{-3}-2.67\times10^{-3}i$ & $183$ & $183$ & $3\times 10^{-13}$ \\
$10^{-6}$ & $180$ & $32$ & $-0.00353$ & $4.23\times10^{-3}-8.87\times10^{-4}i$ & $383$ & $183$ & $9\times 10^{-10}$\\
\hline
\end{tabular}
\caption{Critical Rayleigh numbers for a range of Ekman numbers for differential heating and $\textit{Pr}=1$. The resolutions are given by the maximum spherical harmonic degree $L_\textrm{max}$ and maximum degree of the radial polynomial $N_\textrm{max}$. Also shown are the growth terms $\chi$ computed via the weakly nonlinear analysis.}
\label{tab:eig_results}
 \end{center}
\end{table}

\begin{table}
\begin{center}
\def~{\hphantom{0}}
\begin{tabular}{ l c r r r} \toprule
 $\textit{Ek}$ & Part & $\gamma$ & $\textrm{Real}(\gamma_{AA})$ & $\textrm{Real}(\gamma_{A\bar{A}})$ \\[3pt]
 \hline
 \multirow{3}{3em}{$10^{-3}$}& Total & $1.87\times10^2-1.39\times10^2i$&  $5.26\times10^0$  & $1.82\times10^2$ \\
                    & $\bm{u}$ & $-9.41\times10^0-5.54\times10^0i$ &$-9.83\times10^{-3}$ & $-9.40\times10^0$\\
                             & $T$   & $1.97\times10^2-1.34\times10^2i$& $5.27\times10^0$   & $1.92\times10^2$\\
\hline
\multirow{3}{3em}{$10^{-4}$}& Total & $2.95\times10^3-2.54\times10^3i$&  $-1.56\times10^2$  &$3.10\times10^3$ \\
                    & $\bm{u}$ & $7.34\times10^0+6.35\times10^1i$  & $1.01\times10^2$  & $-9.33\times10^1$\\
                             & $T$   & $2.94\times10^3-2.60\times10^3i$ & $-2.56\times10^2$   &$3.20\times10^3$  \\
\hline
\multirow{3}{3em}{$10^{-5}$}& Total & $1.37\times10^5-6.99\times10^4i$ &  $-4.50\times10^1$  & $1.37\times10^5$\\
                    & $\bm{u}$ & $-1.07\times10^4-6.53\times10^3i$  & $1.36\times10^3$  & $-1.20\times10^4$\\
                             & $T$   & $1.48\times10^5-6.33\times10^4i$ & $-1.41\times10^3$   &  $1.49\times10^5$\\
\hline
\multirow{3}{3em}{$10^{-6}$}& Total & $3.24\times10^{6}-1.49\times10^{6}i$&  $-5.23\times10^{3}$  & $3.25\times10^{6}$\\
                    & $\bm{u}$ & $-3.34\times10^{5}-3.54\times10^{5}i$  &  $9.87\times10^{3}$ & $-3.44\times10^{5}$\\
                             & $T$   & $3.58\times10^{6}-1.14\times10^{6}i$ &  $-1.51\times10^{4}$  & $3.59\times10^{6}$ \\
\hline
\end{tabular}
\caption{The saturation coefficient $\gamma$ in the amplitude equation for varying Ekman number with differential heating and $\textit{Pr}=1$. As well as the total, the contribution due to specific terms is shown.}
\label{tab:coeff_diff}
 \end{center}
\end{table}

With the critical Rayleigh number and wavenumber identified we can now proceed with the weakly nonlinear analysis. The resolvent problems (\ref{equ:qAA}) and (\ref{equ:qAAb}) are solved yielding the second order terms which allow us to compute the saturation coefficient $\gamma$. Using these second order terms the weakly nonlinear coefficients can be computed. As the contribution to $\gamma$ from $\bm{q}_{A\bar{A}}$ (see table \ref{tab:coeff_diff}) far outweighs the contribution from $\bm{q}_{AA}$ it can be concluded that the base flow modification caused by the interaction of the growing eigenmode with its conjugate is the main mechanism through which saturation is obtained. The table further shows that it is primarily the temperature, rather than the velocity field, base-flow modification that is responsible for saturation. {This is not unexpected, the flow acts to modify the mean temperature gradient for saturation.} Recalling the form of $\psi_\epsilon$ we also examine the imaginary part of $\gamma$. The negative value means that $\psi_\epsilon$ will be positive, causing the limit cycle phase to become more positive as the Rayleigh number is increased. In this manner, even though the eigenvector rotates prograde to the planet's rotation, nonlinear effects, primarily a base-flow modification, will cause this rotation to slow down and eventually become retrograde for large enough Rayleigh numbers. This fits with the work of \citet{Feudel_2013} who show that as the Rayleigh number is increased from its critical value the rotation of the thermal Rossby waves slows down and eventually becomes retrograde at this Ekman number {owing to nonlinear effects}.

\begin{figure}
	\centering
	\includegraphics[scale=0.4]{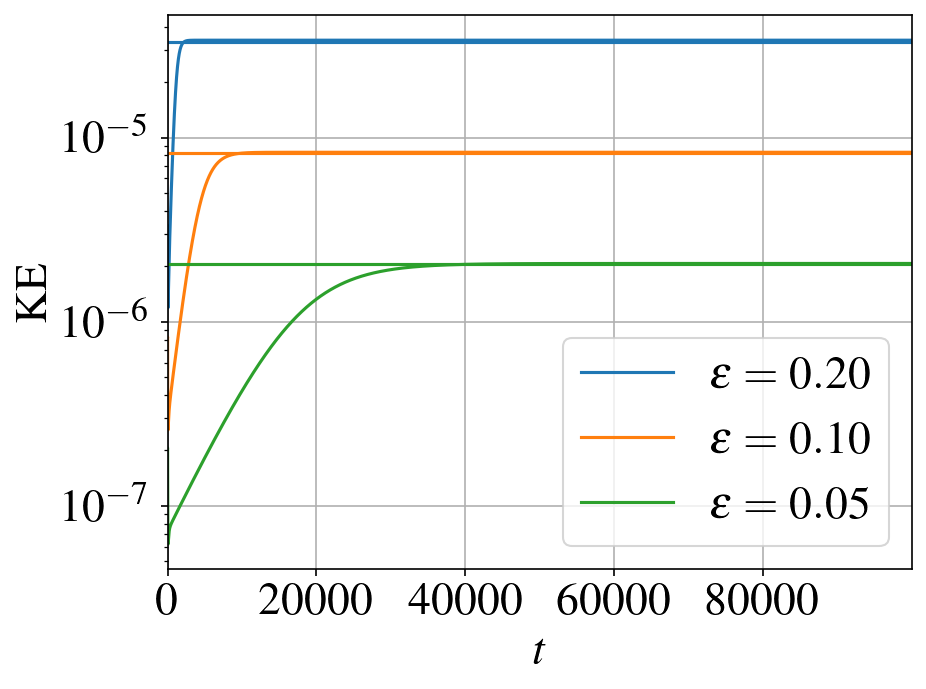}%
    \includegraphics[scale=0.4]{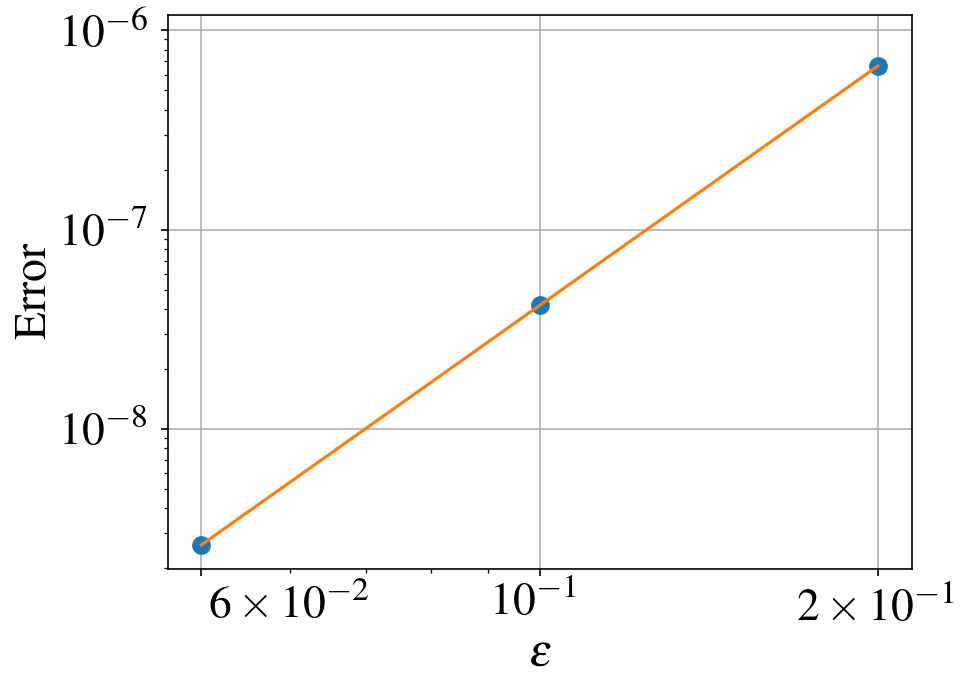}
	\caption{(left) Kinetic energy from fully nonlinear runs for varying values of $\epsilon$. (right) The error in the limit cycle energy as predicted by our weakly nonlinear equations.}
	\label{fig:Ek1e-3verf}
\end{figure}

Before moving on to consider smaller Ekman numbers, we first verify the weakly nonlinear analysis using the method described in \S \ref{sec:maths}. Figure \ref{fig:Ek1e-3verf} (left) shows three nonlinear simulations of the governing equations at different values of $\epsilon$. Also shown as straight horizontal lines are the limit cycle amplitudes predicted by the Stuart--Landau equation. The figure shows that as $\epsilon$ is decreased, i.e. as we move closer to neutrality, the solution takes longer to grow and saturate, with the eventual energy of the limit cycle decreasing. This fits with both the fact that the time scale for the amplitude growth is small depending on $\epsilon$ as $T=\epsilon^2 t$ and that the limit cycle energy depends on $\epsilon$ as given through equation (\ref{equ:energy}). It is also evident that the prediction becomes more accurate as $\epsilon$ is decreased. Figure \ref{fig:Ek1e-3verf} (right) shows the error in our prediction as a function of $\epsilon$. Linear regression shows that the error decreases as $\textrm{Error}\propto \epsilon^\alpha$, where $\alpha\approx3.99$. As this matches with our theoretical prediction of the error being fourth order in $\epsilon$, this verifies our weakly nonlinear procedure.

\begin{figure}
	\centering
	\includegraphics[width=0.33\textwidth]{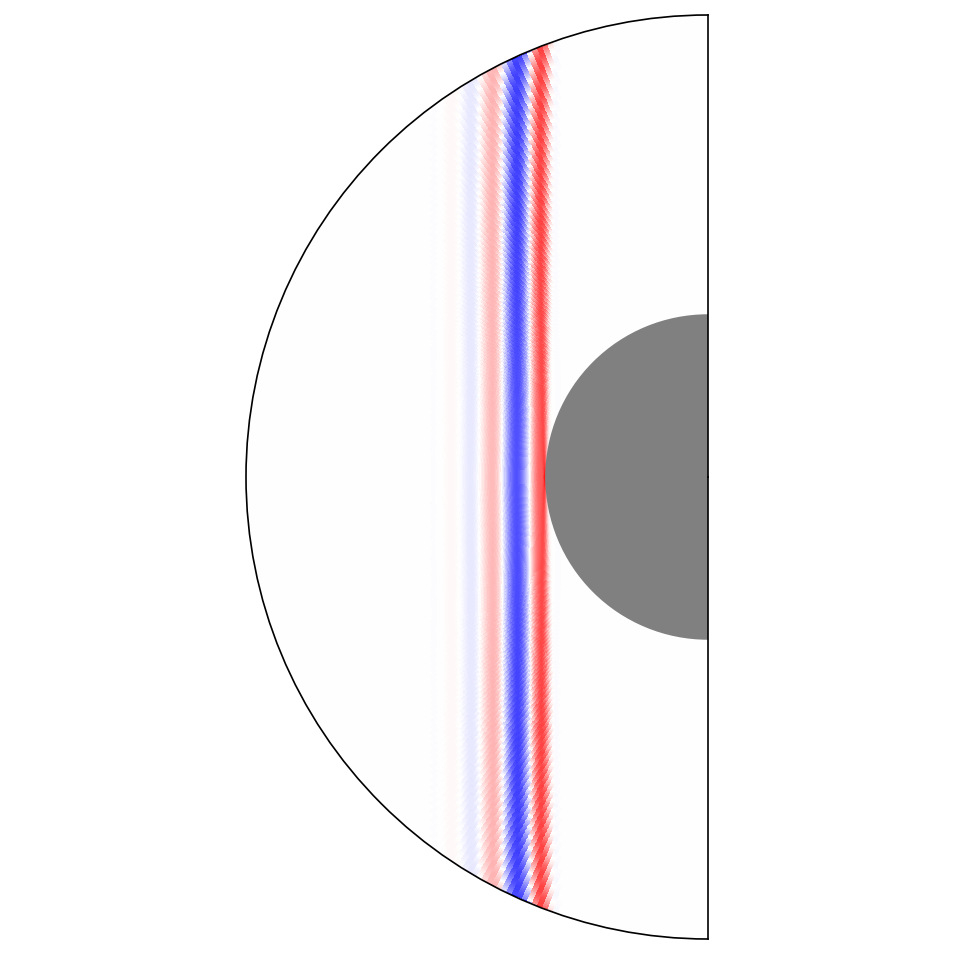}%
    \includegraphics[width=0.33\textwidth]{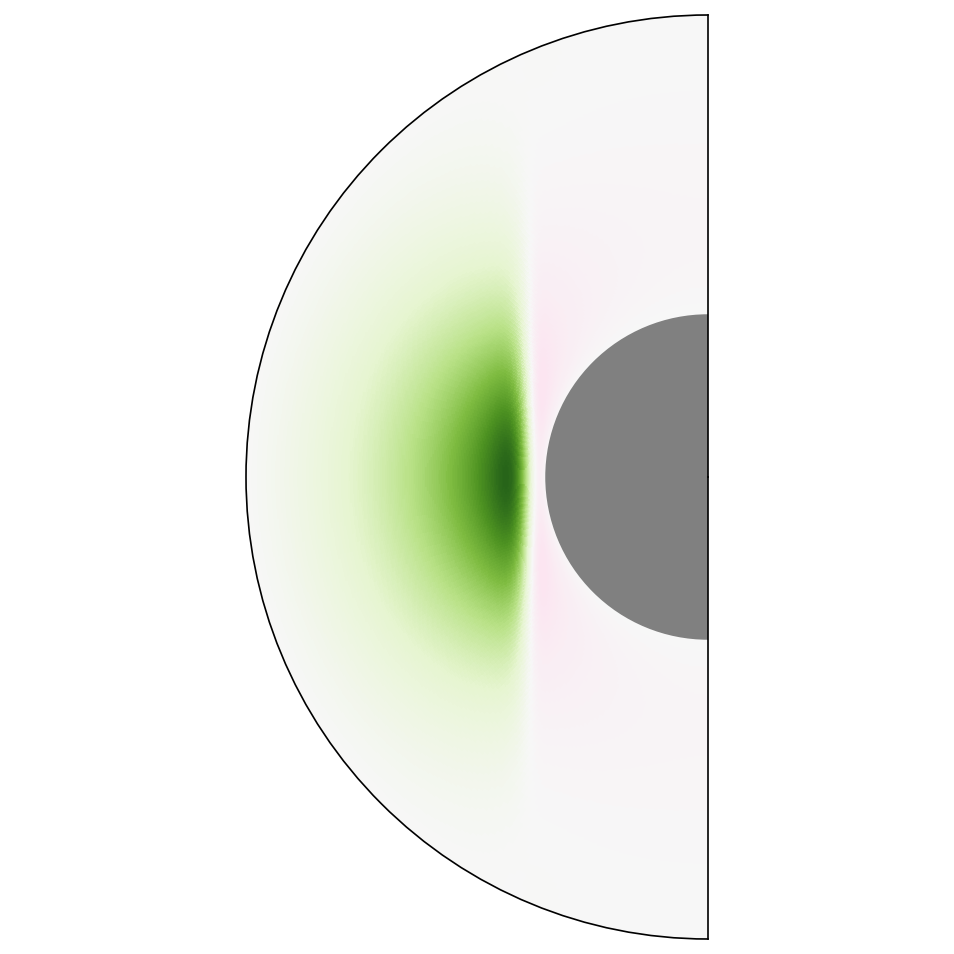}%
    \includegraphics[width=0.33\textwidth]{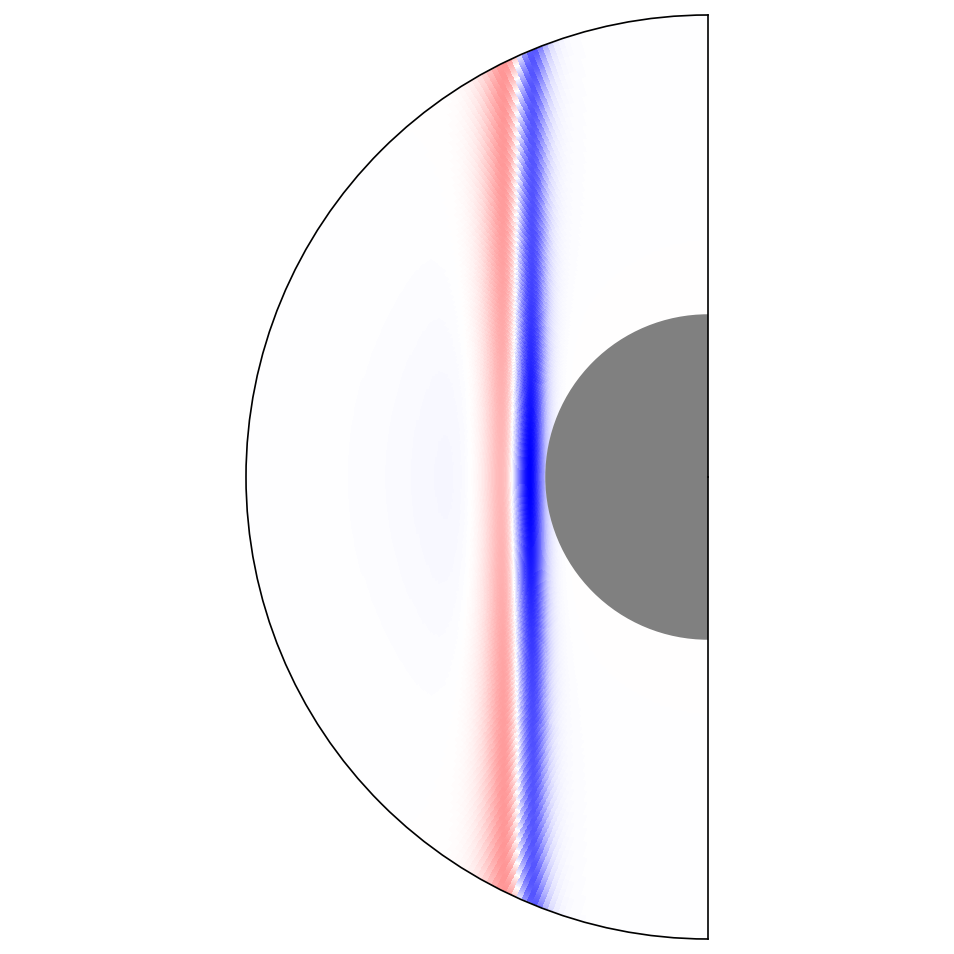}
	\caption{Meridional slices of terms in the weakly nonlinear expansion at $\textit{Ek}=10^{-6}$. From left to right the images display the longitudinal component of $\bm{u}_A$, $T_{A\bar{A}}$, and the longitudinal component of $\bm{u}_{A\bar{A}}$ (red/green is positive and blue/pink is negative).}
	\label{fig:Ek1e-6Rac_meridional}
\end{figure}

\begin{figure}
	\centering
	\includegraphics[width=0.3\textwidth]{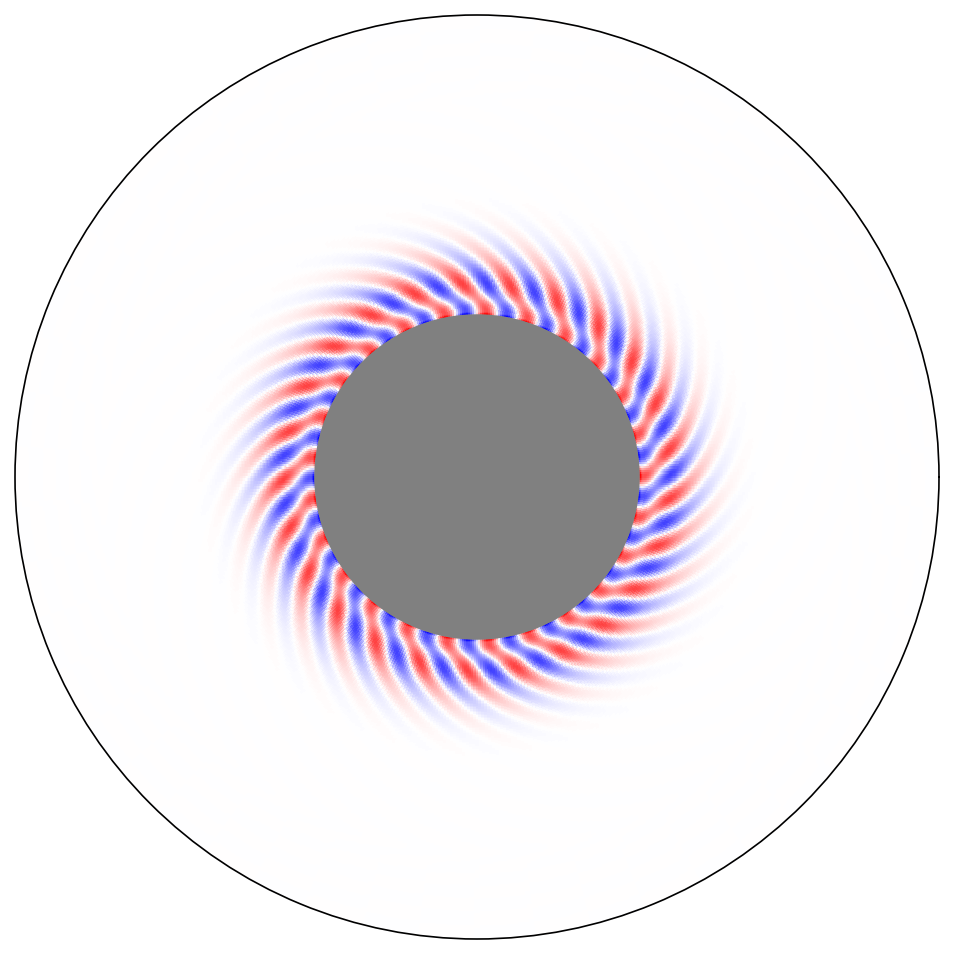}%
    \hspace{0.5cm}%
    \includegraphics[width=0.3\textwidth]{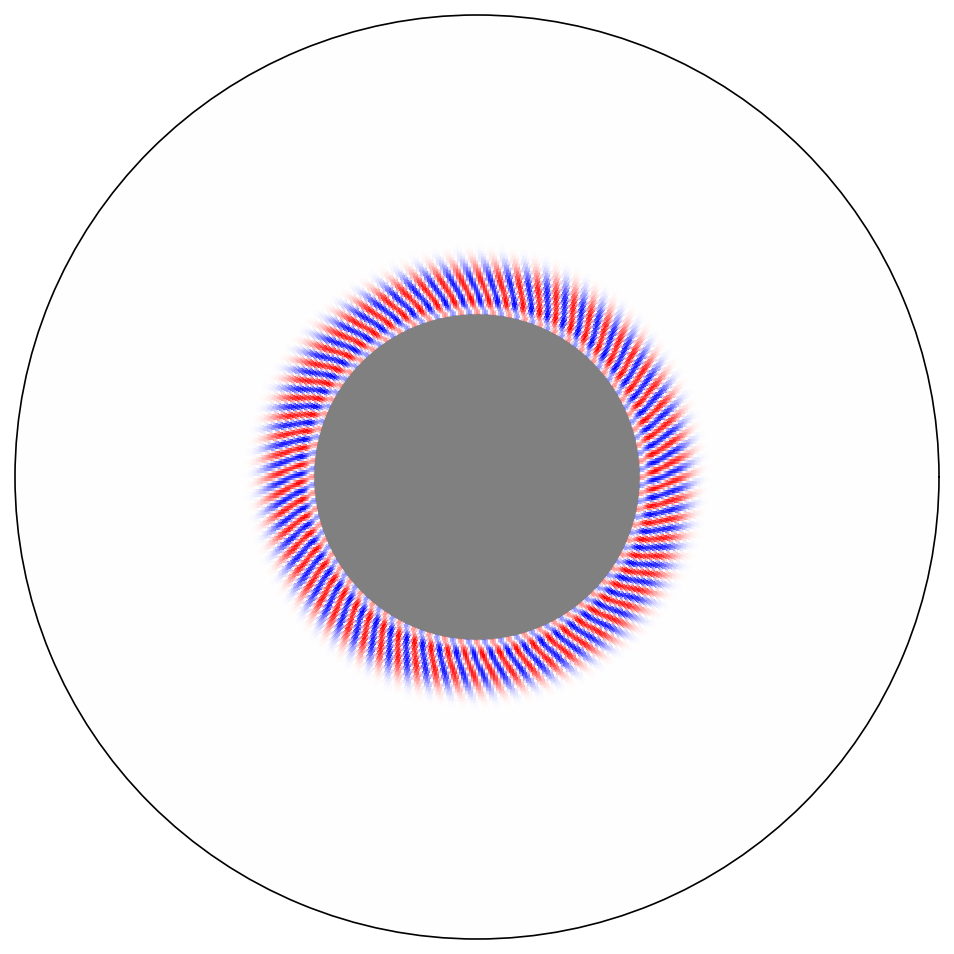}%
    \hspace{0.5cm}%
    \includegraphics[width=0.3\textwidth]{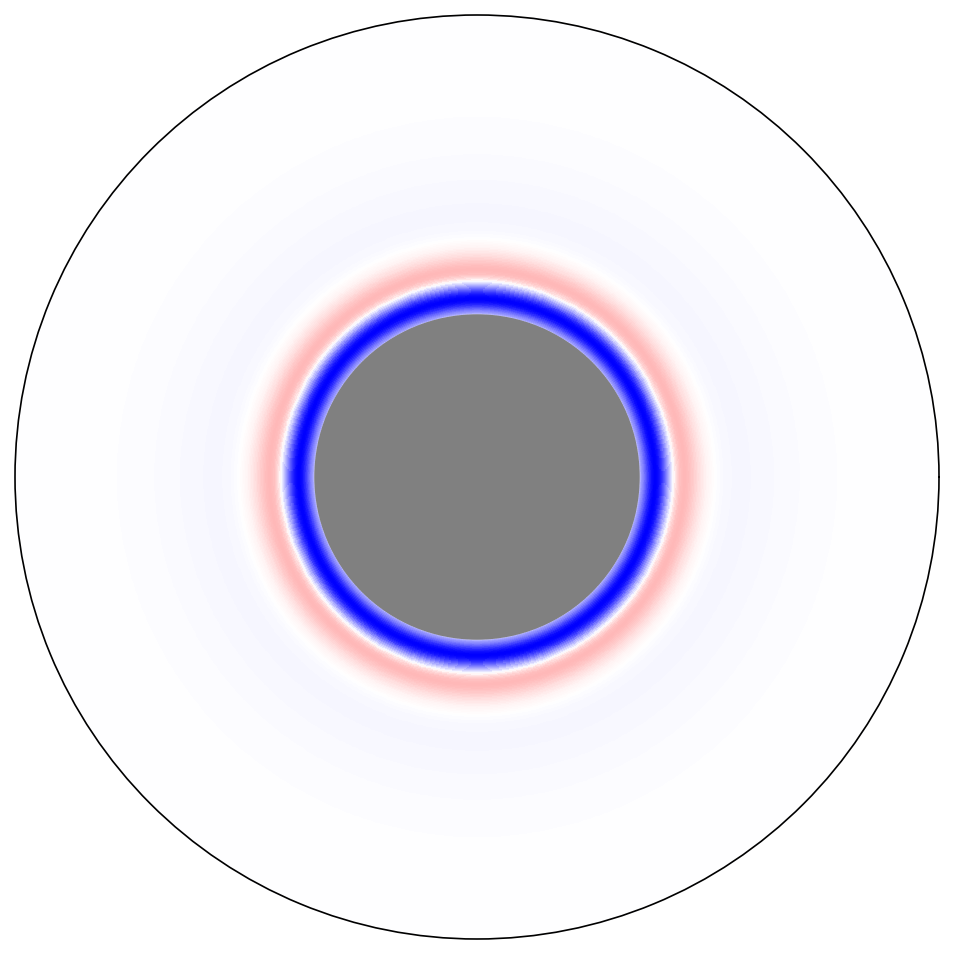}%
	\caption{Equatorial slices of terms in the weakly nonlinear expansion at $\textit{Ek}=10^{-6}$. From left to right the images display the longitudinal component of $\bm{u}_A$,$\bm{u}_{AA}$, and $\bm{u}_{A\bar{A}}$ (red is positive and blue is negative).}
	\label{fig:Ek1e-6Rac_equatorial}
\end{figure}

With the weakly nonlinear analysis described and verified, we now consider smaller values of the Ekman number.  Table \ref{tab:coeff_diff} shows that at all Ekman numbers considered convection onsets supercritically ($\textrm{Real}(\gamma)>0$), with the temperature component of the base flow modification term being primarily responsible. Indeed, the base-flow modification from the velocity promotes {subcriticality}, i.e. makes $\gamma_r$ more negative, as the Ekman number is lowered. Meridional and equatorial slices of the eigenvector and harmonics for $\textit{Ek}=10^{-6}$ are shown in figures \ref{fig:Ek1e-6Rac_meridional} and \ref{fig:Ek1e-6Rac_equatorial}, respectively. As is to be expected from rotating systems by virtue of the Taylor--Proudman theorem, the terms all take the form of alternating columnar structures. The harmonic $\bm{u}_{AA}$ and base-flow modification $\bm{u}_{A\bar{A}}$ show their $m=2m_c$ and $m=0$ dependence, respectively. Similarly to the study of \citet{Guervilly_2016}, the base-flow modification shows a zonal jet structure, albeit confined to a region close to the inner shell boundary due to the differing form of heating used. In fact the longitudinal component is by far the largest component of $\bm{u}_{A\bar{A}}$. In all cases the zonal flow is strongly retrograde near the equator, and has a smaller prograde flow further away which decays to zero so that the zonal jet is confined to the vicinity of the thermal Rossby wave. This form of the zonal flow is to be expected from examining the {dominant} terms of the Reynolds stress \citep{plaut_lebranchu_simitev_busse_2008}. As the Ekman number is lowered the difference between the retrograde and prograde portions of the jet become lessened.

\begin{figure}
	\centering
	\includegraphics[scale=0.4]{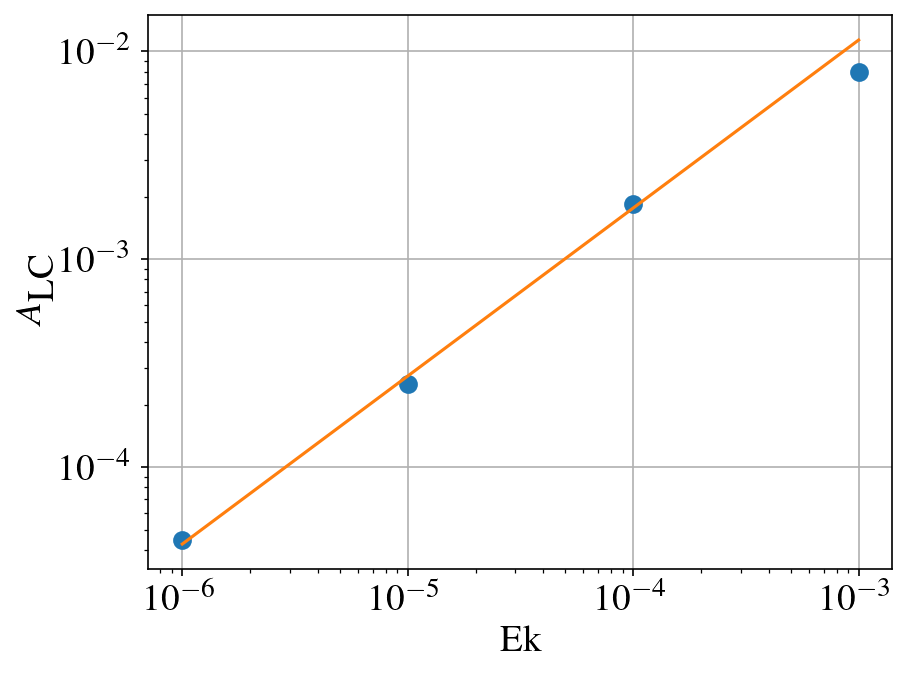}%
    \includegraphics[scale=0.4]{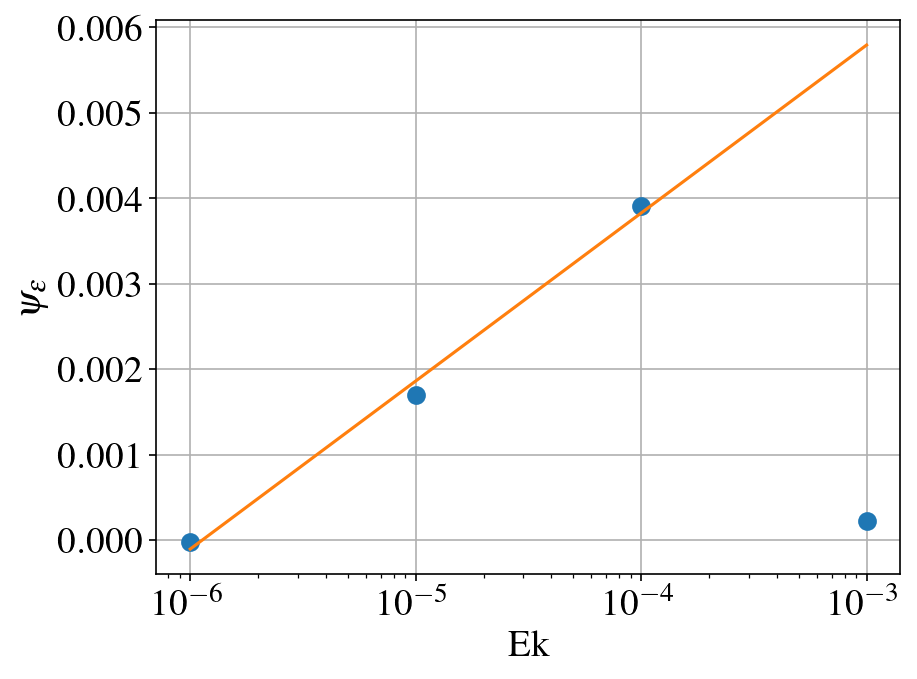}
	\caption{(left) Values of $A_\textrm{LC}$ versus Ekman number (right) Values of $\psi_\epsilon$ versus Ekman number}
	\label{fig:scaling}
\end{figure}

As the Ekman number in the Earth's core is extremely small, around $\textit{Ek}\approx 10^{-15}$, it is computationally infeasible conduct our analysis at realistic planetary values. Therefore, we utilise the results at our computationally tractable Ekman numbers to determine empirical scaling laws for $A_\text{LC}$ and $\psi_\epsilon$ that can be extrapolated. The obtained values of $A_\textrm{LC}$ and $\psi_\epsilon$ are shown in figure \ref{fig:scaling}. Both parameters show a decrease as the Ekman number is lowered, showing definitive scaling laws when $\textit{Ek}=10^{-3}$ is disregarded. The fact that we need to remove this point in order to determine an accurate scaling is not surprising, and is simply a consequence of being far from the asymptotic regime. Linear regression gives the following scalings laws
\begin{gather}
A_\textrm{LC} \propto \textit{Ek}^{0.81}, \\
\psi_\epsilon \approx 8.5\times 10^{-4}\log(\textit{Ek}) + 1.2\times 10^{-2}.\label{equ:psi_scaling}
\end{gather}
It can be hypothesised that as the Ekman number is decreased even further that the scaling for the amplitude will ultimately become $A_\textrm{LC} \propto \textit{Ek}^{4/5}$. Combining this scaling with previous work \citep{dormy_2004,Barik2023} would then imply that $A_\textrm{LC} \propto \textit{Ra}_c^{-1}$. The result is that as the Ekman number is lowered the limit cycle produced near the onset of convection will have decreasing energy. {Moreover,} the scaling for $\psi_\epsilon$ given by (\ref{equ:psi_scaling}) shows that as the Ekman number is lowered, the phase change caused through nonlinear effects will decrease until it becomes negative. Hence, there is a critical Ekman number where the phase change switches from being a prograde rotation to a retrograde one. This can be seen on figure \ref{fig:scaling} (right) to lie close to $\textit{Ek}=10^{-6}$. Extrapolating this further shows that at planetary values we should expect nonlinearities to reinforce the prograde rotation of the neutral mode, producing a limit cycle rotating prograde to the planet.

\subsection{Internal heating with varying Prandtl number}\label{sec:internal_heating}

With the weakly nonlinear procedure now illustrated, we turn our attention to searching for subcritical convection in a shell. Based on previous works \citep{soward_1977,plaut_lebranchu_simitev_busse_2008,Guervilly_2016,Kaplan2017} we will carry out this search using {a system that is internally heated}. Given the similarities between the asymptotic theory of the onset of convection for spheres and shells for Earth-like radius ratios{, as well as the study of \citet{plaut_lebranchu_simitev_busse_2008}, we might expect that for internal heating a spherical shell can onset subcritically given that it certainly can in a full sphere \citep{Guervilly_2016,Kaplan2017} as well as in a quasi-geostrophic model of a rotating shell \citep{plaut_lebranchu_simitev_busse_2008}}. Nevertheless, we stress that the advantage of our current numerical weakly nonlinear procedure is that the type of bifurcation can be determined {efficiently} by performing one extra eigenvalue problem solve, and two linear boundary value problem solves for the harmonic and base-flow modification.

\begin{figure}
	\centering
	\includegraphics[width=0.33\textwidth]{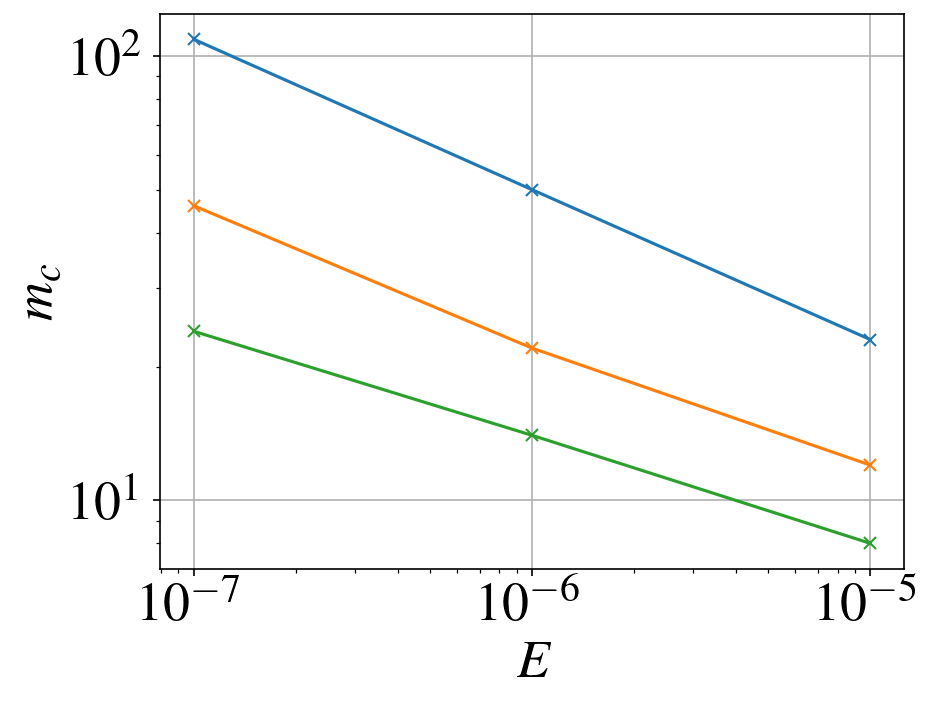}%
    \includegraphics[width=0.33\textwidth]{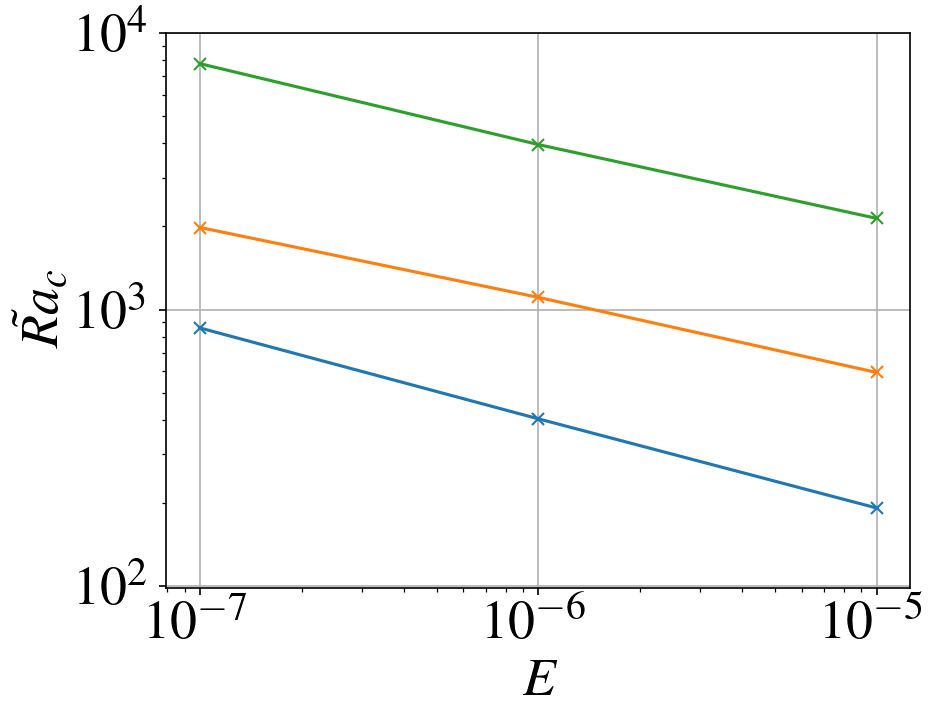}%
    \includegraphics[width=0.33\textwidth]{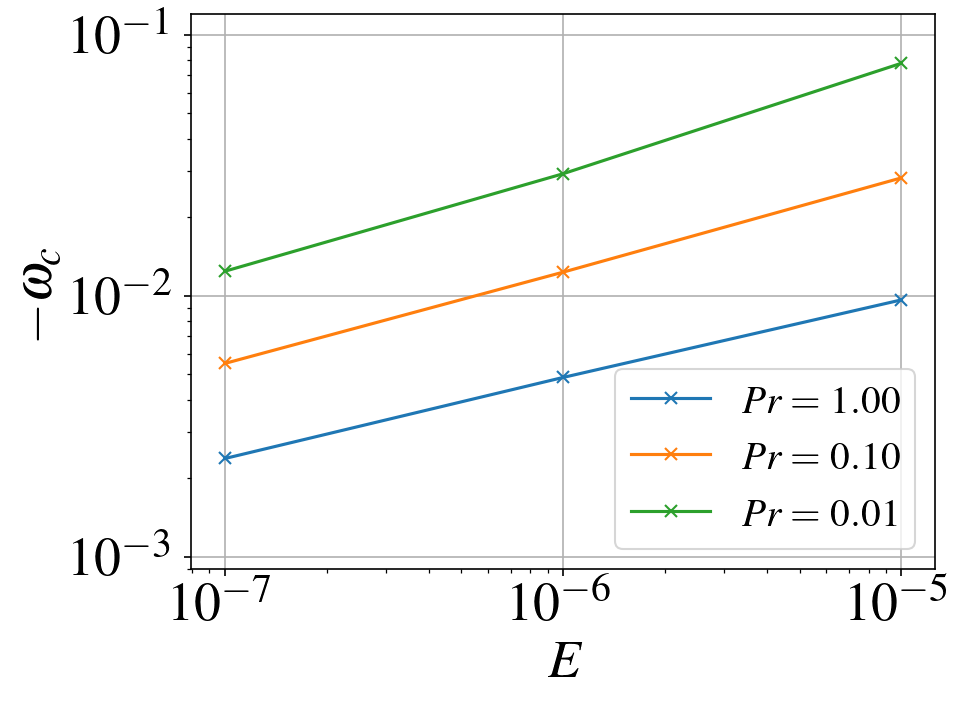}%
	\caption{Dependence of $m_c$(left), $\tilde{\textit{Ra}}_c$ (middle), and $\omega_c$ (right) on the Ekman number for internal heating.}
	\label{fig:internal_Params}
\end{figure}

\begin{table}
\begin{center}
\def~{\hphantom{0}}
\begin{tabular}{l l c c c c c c c} \toprule
$\textit{Pr}$ & $\textit{Ek}$ & $\tilde{\textit{Ra}}_c$ & $m_c$ & $\omega_c$ & $\chi$ & $L_\textrm{max}$ & $N_\textrm{max}$ & $\mathcal{R}$\\[3pt]
\hline
 \multirow{3}{1em}{$1$} &$10^{-5}$ & 192 & 23 & -0.00964 & $1.04\times10^{-2}-4.32\times10^{-3}i$ & 183 & 183& $1\times10^{-14}$ \\
&$10^{-6}$ & 403  & 50 & -0.00486 & $4.97\times10^{-3}-1.97\times10^{-3}i$ & 383 &183 & $6\times10^{-10}$ \\
&$10^{-7}$ & 859 &  109& -0.00238 & $2.35\times10^{-3}-8.91\times10^{-4}i$ &  383& 383& $3\times10^{-12}$ \\
\hline
 \multirow{3}{1em}{$0.1$} &$10^{-5}$ &593 & 12 & -0.0283 & $1.57\times10^{-2}-3.54\times10^{-4}i$ &  183& 183&  $1\times10^{-14}$\\
&$10^{-6}$ & 1108 & 22 & -0.0123 &  $5.75\times10^{-3}+5.40\times10^{-4}i$&  383& 183& $2\times10^{-10}$ \\
&$10^{-7}$ & 1982 & 46 & -0.00550 & $2.11\times10^{-3}+5.56\times10^{-4}i$ &  383& 383& $5\times10^{-14}$ \\
\hline
 \multirow{3}{1em}{$0.01$} &$10^{-5}$ & 2140 & 8 & -0.0778 & $1.75\times10^{-2}+1.21\times10^{-2}i$ & 183 &183 & $2\times10^{-14}$ \\
&$10^{-6}$ & 3955 & 14 & -0.0293 & $5.94\times10^{-3}+2.14\times10^{-3}i$ & 383 & 183& $5\times10^{-13}$ \\
&$10^{-7}$ & 7752 & 24 & -0.0124 & $2.37\times10^{-3}+8.90\times10^{-4}i$ & 383 & 383& $5\times10^{-12}$ \\
\hline
\end{tabular}
\caption{Similar to table \ref{tab:eig_results} but for internal heating}
 \end{center}
\end{table}

In order to map out the parameter regime in which subcritical convection is possible in a shell, we consider $\textit{Ek}\in\{10^{-5},10^{-6},10^{-7}\}$ and $\textit{Pr}\in\{1,0.1,0.01\}$. This parameter region is motivated by the case of a sphere \citep{Guervilly_2016,Kaplan2017} with the additional case of $\textit{Pr}=1$ added. Figure \ref{fig:internal_Params} shows the critical wavenumbers, Rayleigh numbers, and rotational frequencies obtained for these parameters. In all cases a clear Prandtl-number-dependent scaling law with Ekman number is found. As the Prandtl number is decreased the critical azimuthal wavenumbers decrease, and critical Rayleigh numbers increase. Consequently, it is more difficult to get to the asymptotic regime. 

\begin{figure}
	\centering
	\includegraphics[width=0.33\textwidth]{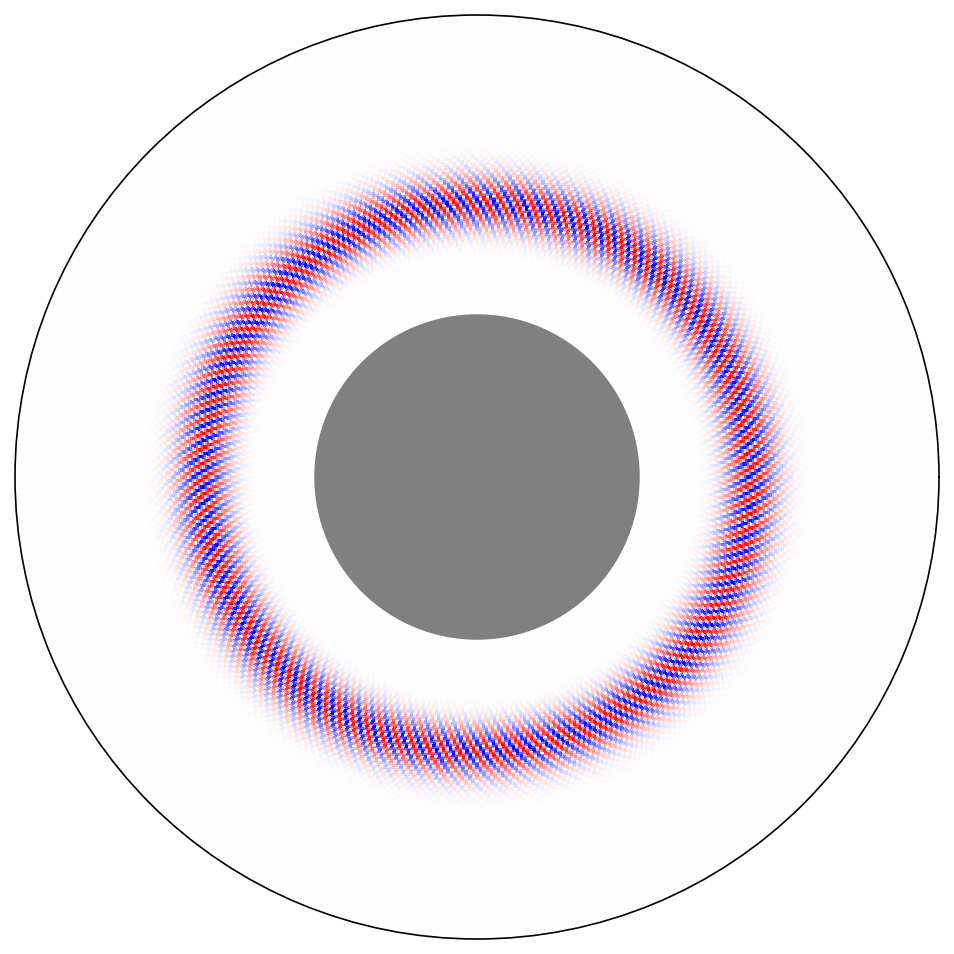}%
    \includegraphics[width=0.33\textwidth]{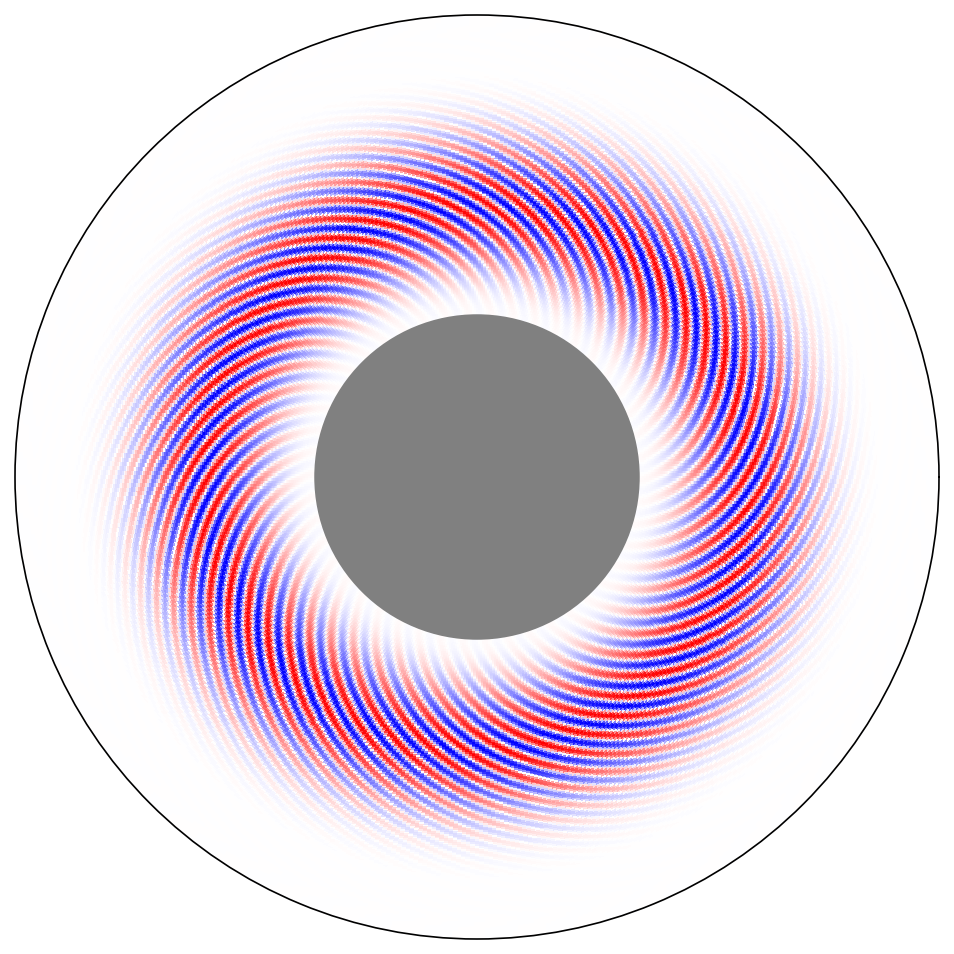}%
    \includegraphics[width=0.33\textwidth]{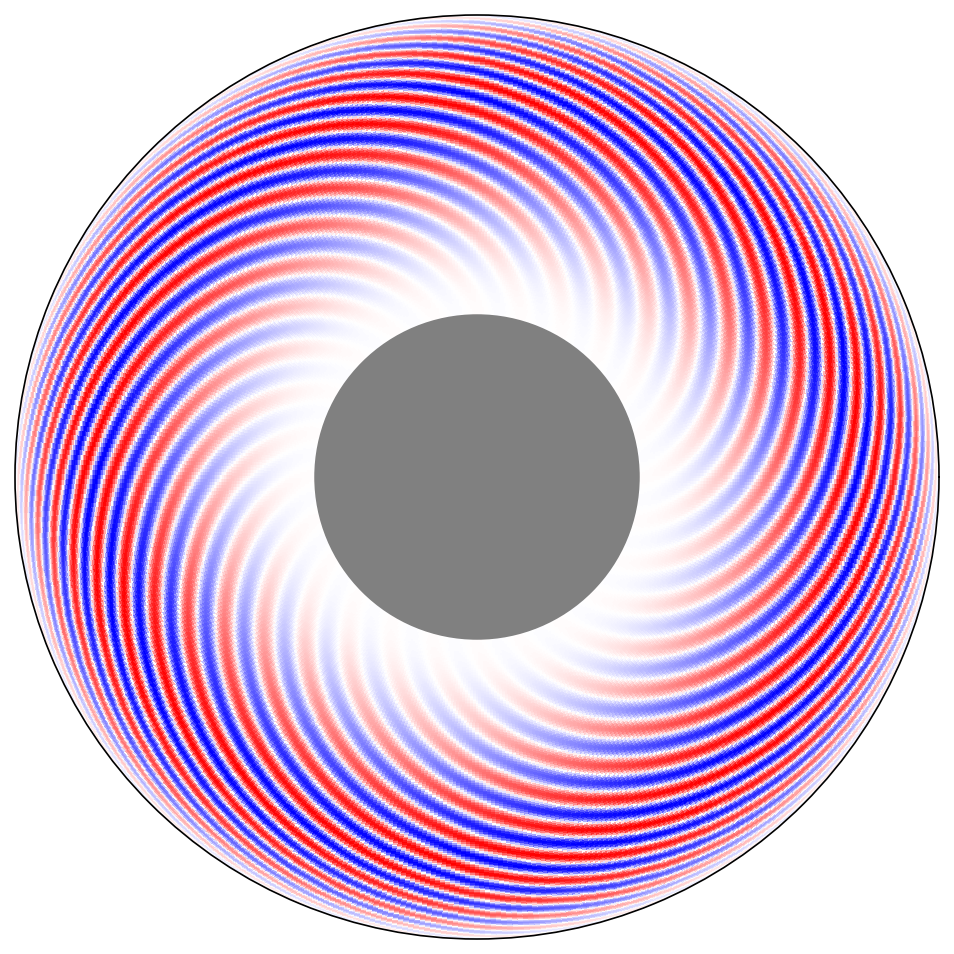}%
	\caption{Equitorial slices of the longitudinal component of $\bm{u}_A$ at $\textit{Ek}=10^{-7}$ for $\textit{Pr}=1$ (left), $\textit{Pr}=0.1$ (middle), and $\textit{Pr}=0.01$ (right) (red is positive and blue is negative).}
    \label{fig{equ:int}}
\end{figure}

\begin{figure}
	\centering
	\includegraphics[width=0.33\textwidth]{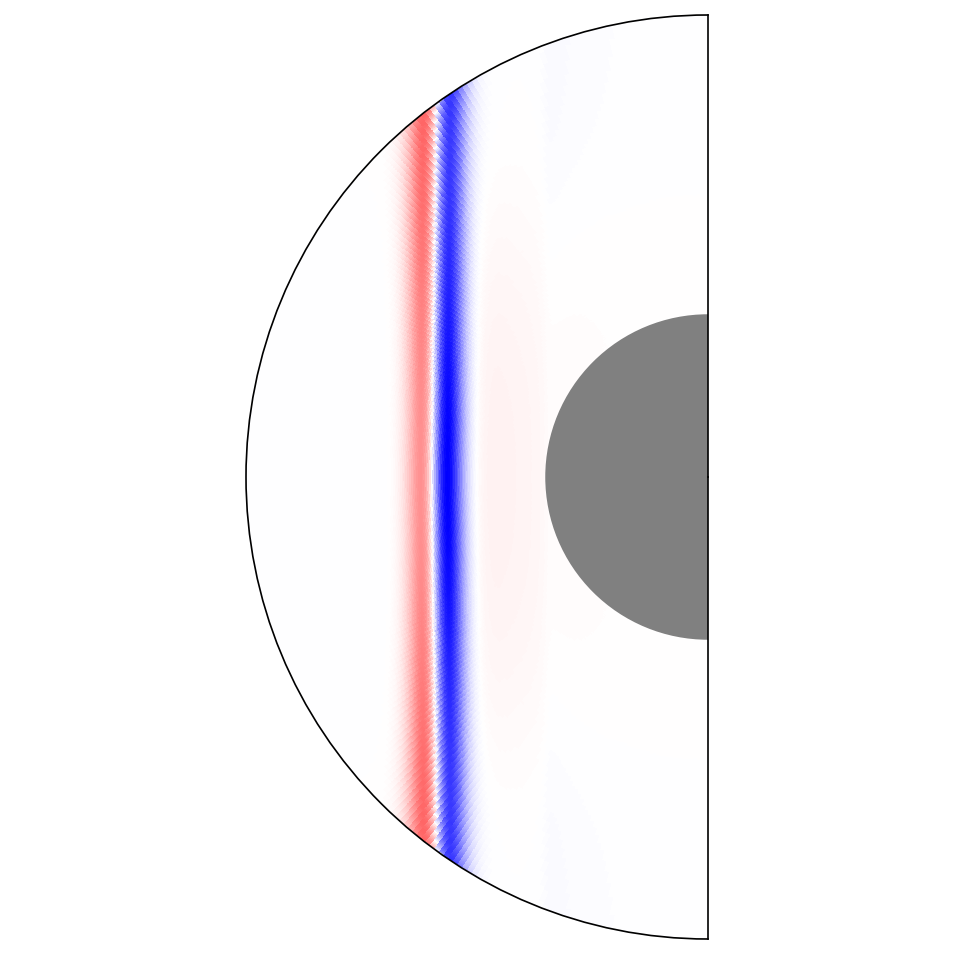}%
    \includegraphics[width=0.33\textwidth]{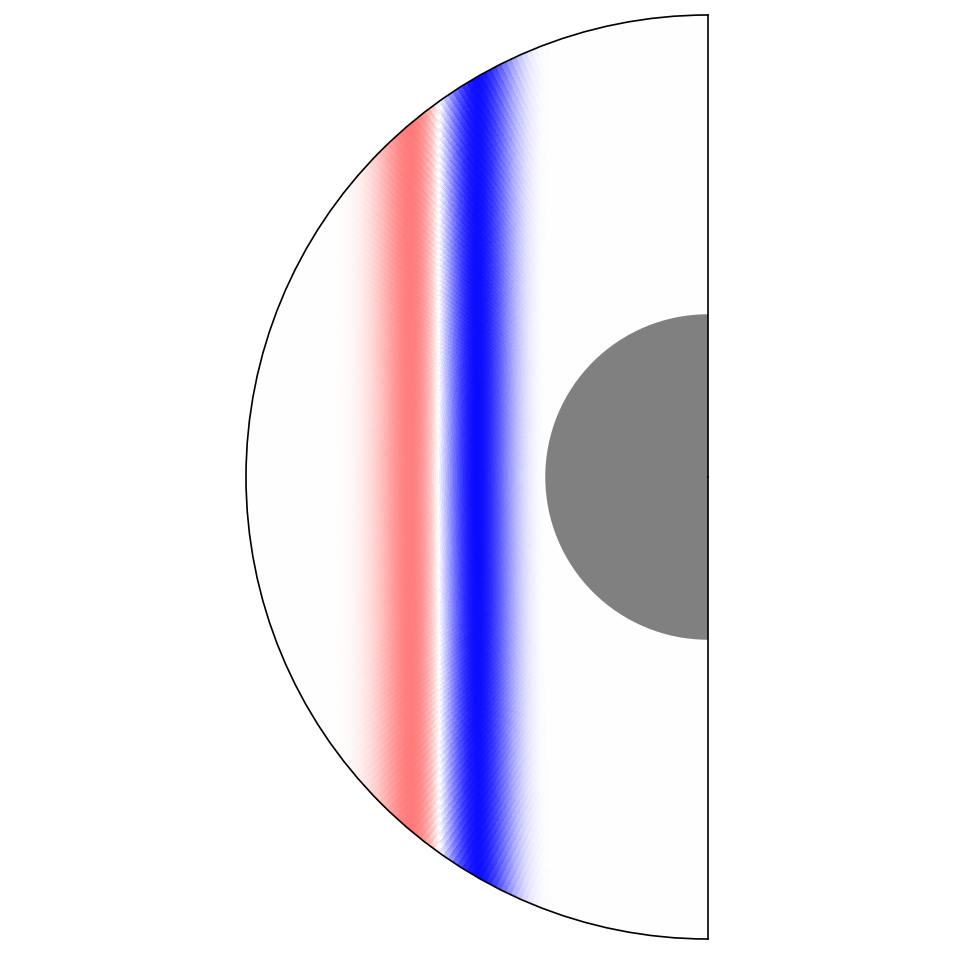}%
    \includegraphics[width=0.33\textwidth]{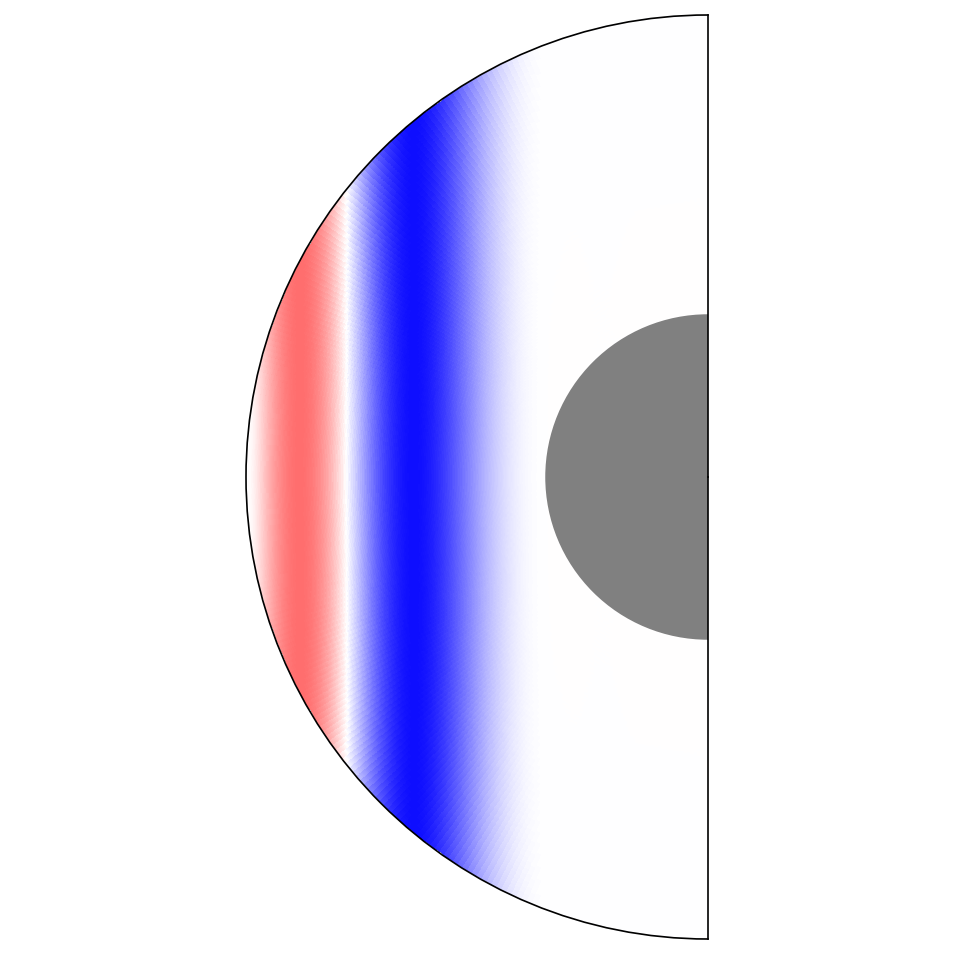}
    \includegraphics[width=0.33\textwidth]{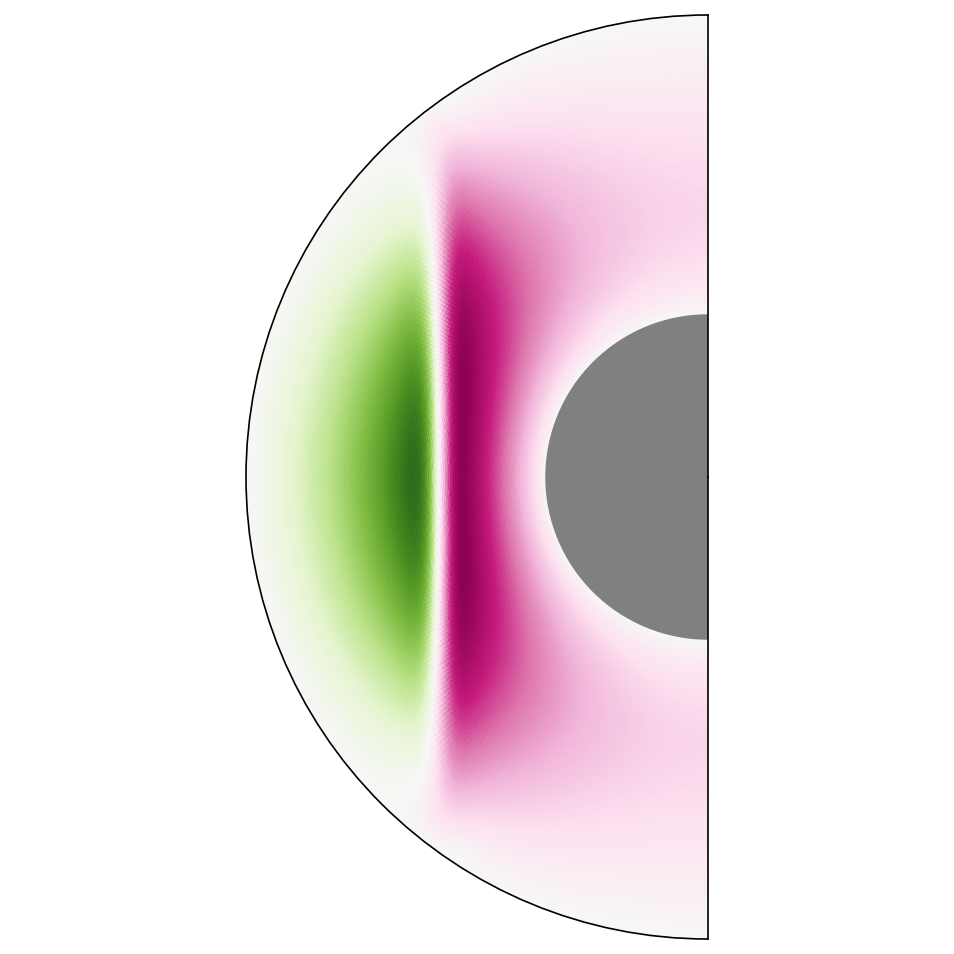}%
    \includegraphics[width=0.33\textwidth]{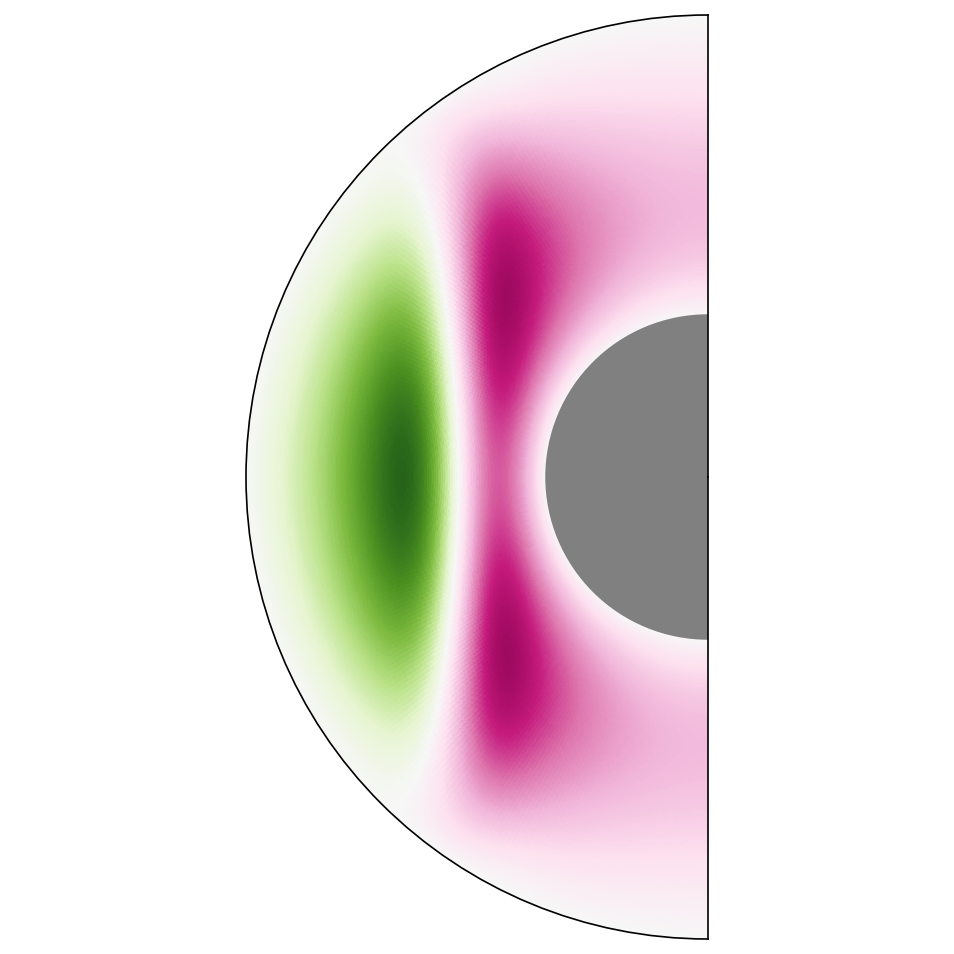}%
    \includegraphics[width=0.33\textwidth]{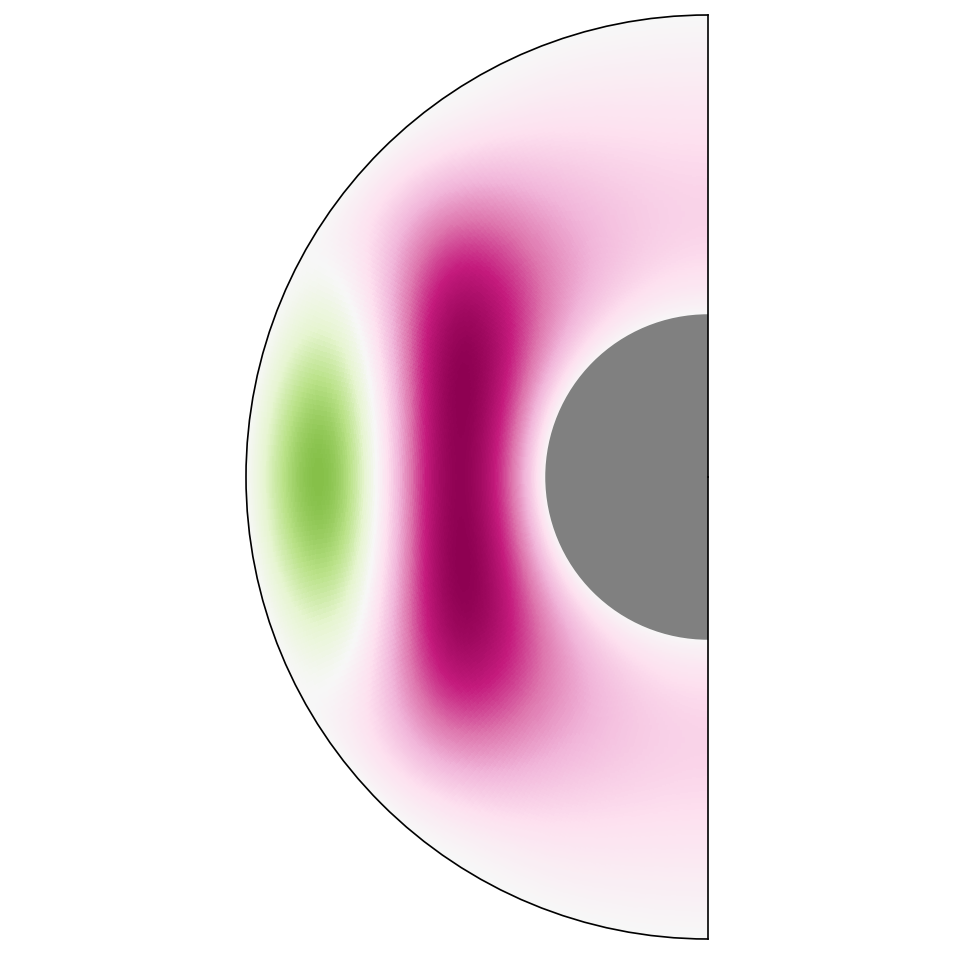}
	\caption{Meridional slices of the longitudinal component of $\bm{u}_{A\bar{A}}$ (top row), and $T_{A\bar{A}}$ (bottom row) with internal heating at $\textit{Ek}=10^{-7}$ for $\textit{Pr}=1$ (left), $\textit{Pr}=0.1$ (middle), and $\textit{Pr}=0.01$ (right) (red/green is positive and blue/pink is negative).}
    \label{fig:merid_int}
\end{figure}

With internal heating, convection still onsets as rotation-axis-aligned columnar structures, with figure \ref{fig:merid_int} showing equatorial slices of the unstable mode for our three considered Prandtl numbers. In all cases, convection now onsets in the middle of the shell, as predicted by asymptotic theory \citep{dormy_2004}. We note that due to the low radius ratio, this is similar to the case of the full sphere \citep{Jones_2000}. As the Prandtl number is lowered convection still onsets at a critical cylindrical radius in the interior of the shell, however, it now extends radially outwards towards the outer shell radius.

\begin{table}
  \begin{center}
\def~{\hphantom{0}}
\begin{tabular}{l l c r r r} \toprule
 $\textit{Pr}$ & $\textit{Ek}$ & Part & $\gamma$ & $\textrm{Real}(\gamma_{AA})$ & $\textrm{Real}(\gamma_{A\bar{A}})$ \\[3pt]
 \hline
 \multirow{9}{2em}{$1$} & \multirow{3}{2.25em}{$10^{-5}$}& Total & $1.29\times 10^{5}-1.01\times 10^{5}i$&  $1.57\times 10^{3}$  & $1.27\times 10^{5}$\\
                 &   & $\bm{u}$ & $-1.53\times 10^{4} -1.86\times 10^{4}i$ & $6.71\times 10^{2}$ & $-1.60\times 10^{4}$\\
                 &            & $T$   & $1.44\times 10^{5}-8.19\times 10^{4}i$ &  $8.96\times 10^{2}$   & $1.43\times 10^{5}$ \\
                 & \multirow{3}{2.25em}{$10^{-6}$}& Total & $2.68\times10^{6}-7.39\times10^{6}i$& $4.11\times10^{4}$   & $2.64\times10^{6}$ \\
                 &   & $\bm{u}$ & $-1.47\times10^6-7.35\times10^5i$& $1.72\times10^4$ & $-1.48\times10^6$\\
                 &            & $T$   &$4.14\times10^6-6.66\times10^6i$ & $2.39\times10^4$     & $4.12\times10^6$ \\
                 & \multirow{3}{2.25em}{$10^{-7}$}& Total & $-9.10\times10^8-2.16\times10^8i$ & $2.10\times10^6$   & $-9.12\times10^8$ \\
                 &   & $\bm{u}$ & $-1.03\times10^8+2.21\times10^8i$ & $4.89\times10^5$ & $-1.04\times10^8$\\
                 &            & $T$   & $-8.07\times10^8-4.37\times10^8i$  & $1.61\times10^6$    & $-8.09\times10^8$ \\
 \hline
\multirow{9}{2em}{$0.1$} & \multirow{3}{2em}{$10^{-5}$}& Total & $1.02\times10^3-2.60\times10^4i$& $-3.12\times10^2$   & $1.33\times10^3$ \\
                 &   & $\bm{u}$ & $-5.54\times10^3-1.26\times10^4i$&$-2.52\times10^2$  & $-5.29\times10^3$\\
                 &            & $T$   & $6.56\times10^3-1.35\times10^4i$  &  $-5.98\times10^1$   & $6.62\times10^3$ \\
                 & \multirow{3}{2em}{$10^{-6}$}& Total &$-9.55\times10^4-3.66\times10^4i$ &  $1.17\times10^2$   & $-9.56\times10^4$\\
                
                 &   & $\bm{u}$ & $-6.30\times10^4+2.05\times10^2i$ & $5.13\times10^1$& $-6.31\times10^4$ \\
                 &            & $T$   &$-3.25\times10^4-3.68\times10^4i$ & $6.53\times10^1$     &  $-3.25\times10^4$\\
                 & \multirow{3}{2em}{$10^{-7}$}& Total &$-2.14\times10^5+1.95\times10^4i$ & $-2.40\times10^0$   & $-2.14\times10^5$ \\
                 &   & $\bm{u}$ & $-1.16\times10^5+6.36\times10^4i$ & $-9.71\times10^0$ & $-1.16\times10^5$\\
                 &            & $T$   & $-9.82\times10^4-4.40\times10^4i$ & $7.31\times10^0$    & $-9.83\times10^4$ \\
 \hline
\multirow{9}{2em}{$0.01$} & \multirow{3}{2em}{$10^{-5}$}& Total &$1.16\times10^3-1.16\times10^3i$ &  $7.84\times10^1$  &$1.08\times10^3$ \\
                 &   & $\bm{u}$ & $5.76\times10^2-1.04\times10^3i$  & $7.82\times10^1$ & $4.98\times10^2$\\
                 &            & $T$   & $5.80\times10^2-1.19\times10^2i$ & $2.11\times10^{-1}$    & $5.80\times10^2$ \\
                 & \multirow{3}{2em}{$10^{-6}$}& Total & $2.92\times10^1-1.07\times10^4i$& $4.02\times10^2$   & $-3.73\times10^2$\\
                 &   & $\bm{u}$ & $-1.09\times10^3-7.83\times10^3i$ & $3.42\times10^2$ & $-1.43\times10^3$\\
                 &            & $T$   &$1.12\times10^3-2.83\times10^3i$ &   $6.03\times10^1$     & $1.06\times10^3$ \\
                 & \multirow{3}{2em}{$10^{-7}$}& Total &$-3.00\times10^3-2.03\times10^4i$ & $1.36\times10^3$ & $-4.36\times10^3$ \\
                 &   & $\bm{u}$ & $-4.02\times10^3-1.55\times10^4i$ & $1.19\times10^3$ & $-5.22\times10^3$\\
                 &            & $T$   & $1.02\times10^3-4.83\times10^3i$ & $1.70\times10^2$    & $8.55\times10^2$ \\
\hline
\end{tabular}
\caption{The saturation coefficient $\gamma$ in the amplitude equation for varying Ekman number and Prandtl number for internal heating. As well as the total, the contribution due to specific terms is shown.}
\label{tab:coeff_int}
 \end{center}
\end{table}

\begin{figure}
	\centering
	\includegraphics[scale=0.6]{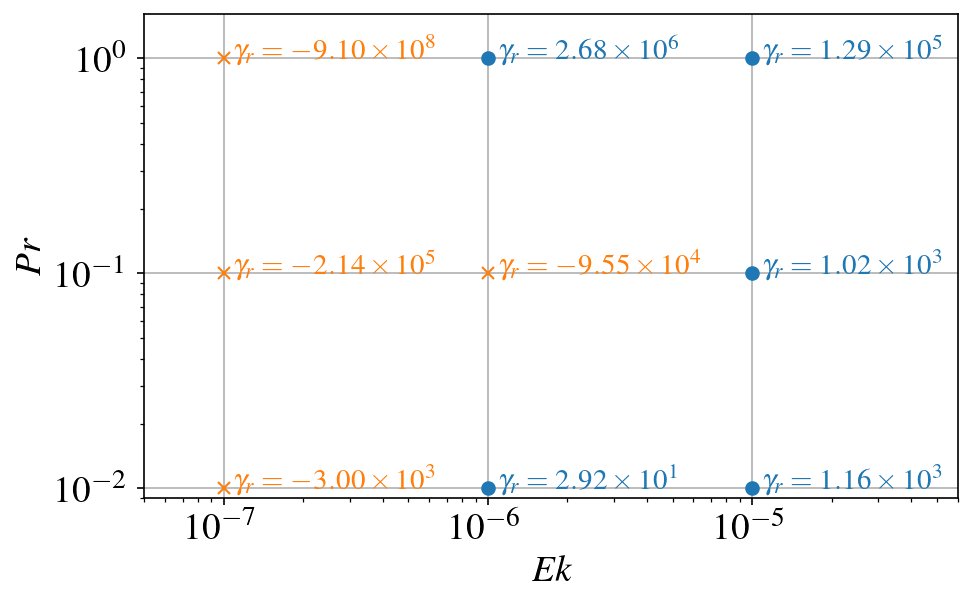}
	\caption{Regime diagram showing the nature of the bifurcations obtained for internal heating. Supercritical Hopf bifurcations are marked with blue circles and subcritical Hopf bifurcations are marked with orange crosses.}
    \label{fig:regime}
\end{figure}

Table \ref{tab:coeff_int} shows the weakly nonlinear coefficient $\gamma$ for {cases of} internal heating. For each of the three Prandtl numbers considered, convection onsets subcritically for small enough Ekman number. The regions of subcriticality are highlighted in the regime diagram (figure \ref{fig:regime}). Our regime diagram is in agreement with the {full sphere} results of \citet{Guervilly_2016} for the overlapping cases of $\textit{Pr}=0.1$ and $\textit{Pr}=0.01$, showing that, as was to perhaps be expected from the asymptotic theory \citep{dormy_2004}, a low radius ratio shell behaves similarly to a full sphere. In comparing our results with that of \citet{Guervilly_2016} we have taken the oscillating case on their regime diagram to be subcritical. In other words, as long as nonlinearities are not saturating the instability to a stable limit cycle near the critical Rayleigh number, we are taking the behaviour to be subcritical. Additionally our regime diagram shows that for {internal heating and $\textit{Pr}=1$, subcritical convection can occur for small enough Ekman number. This agrees with the weakly nonlinear analysis of \citet{plaut_lebranchu_simitev_busse_2008}, where subcritical convection occurs for a quasi-geostrophic model of rotating convection in a shell at $\textit{Pr}=1$ for low Ekman numbers.}

By examining table \ref{tab:coeff_int}, it is evident that, similarly to differential heating, the base-flow modification is the dominant effect determining the saturation mechanism. For unity Prandtl number, the temperature component of the base-flow modification is still the predominant effect but the difference between this and the velocity component is less than in the case of differential heating. Hence, the change to a subcritical bifurcation at $\textit{Ek}=10^{-7}$ occurs when the temperature component of the base-flow modification changes to a subcritical rather than supercritical term.  For lower Prandtl numbers, {i.e. $\textit{Pr}\leq0.1$}, the velocity component of the base-flow modification becomes the dominant term determining the type of bifurcation. Whilst for $\textit{Pr}=0.1$ both the temperature and velocity field base-flow terms promote instability for small enough Ekman numbers, for $\textit{Pr}=0.01$ it is the velocity field alone that causes a subcritical bifurcation.

Figure \ref{fig:merid_int} shows the zonal flow contained in the base-flow modification. As for differential heating, the zonal flow is the dominant component of this term. In all cases the zonal flow transitions from retrograde to prograde, and is concentrated near the critical radius where convection onsets. As the Prandtl number is lowered the zonal flow spreads out for fixed Ekman number as a consequence of being further from the asymptotic regime. However, for each Prandtl number lowering the Ekman number causes the retrograde part of the jet to become closer in magnitude to its prograde counterpart. This behaviour is consistent with that seen for differential heating.

Because the zonal jet that arises in response to the instability always promotes {subcriticality} for low enough Ekman numbers (mirroring the case of differential heating), the change to subcritical behaviour overall can mainly be attributed to the temperature component of the base-flow modification. For differential heating the temperature component is always dominant over the velocity component, causing supercritical behaviour at all Ekman numbers {considered}. For internal heating we obtain subcritical behaviour in two main ways. The first, seen for $\textit{Pr}=1$ and $\textit{Pr}=0.1$, is that the temperature component switches to have a subcritical effect for low enough Ekman numbers. The second, {observed for $\textit{Pr}=0.1$ and $\textit{Pr}=0.01$, is that the velocity component becomes the dominant term. This can be attributed to heat transport being less efficient at smaller Prandtl numbers, and allows convection to onset subcritically even when the temperature base-flow modification opposes it (as seen for $\textit{Pr}=0.01$).} It appears from the table that for $\textit{Pr}=0.01$ the temperature contribution has started to switch from positive to negative, indicating that the first effect is probably an asymptotic effect that occurs at low Ekman numbers. We hypothesise that at low enough Ekman numbers both terms will become {subcriticality}-promoting terms for $\textit{Pr}=0.01$. The reason that $\textit{Pr}=0.1$ becomes subcritical at the largest Ekman number can then be attributed to it having the right balance between the speed at which it approaches the asymptotic regime, and the relative importance of the velocity and temperature components of the base-flow. 

The fact that the zonal flow always promotes subcriticality for low enough Ekman numbers can be understood as the zonal flow locally reducing the rotation rate, easing the constraining effect of rapid rotation on convection and allowing it to set in at lower Rayleigh numbers. However, the role of the temperature base-flow modification is trickier to determine. For differential heating, the temperature base-flow modification is localised to the location where the Taylor columns arise (figure \ref{fig:Ek1e-6Rac_meridional}). As it locally reduces the temperature gradient, it suppresses the instability leading to a supercritical bifurcation. For internal heating, figure \ref{fig:merid_int} shows that the form the temperature base-flow modification is less localised. Even though it locally reduces the temperature gradient near the critical radius where convection onsets, it also raises the temperature gradient away from this point which may promote instability leading to the subcritical behaviour observed for internal heating at some parameters.  Although it is hard to pinpoint exactly why the temperature base-flow modification promotes instability for internal heating at certain parameters, it is important to note that the weakly nonlinear analysis systematically identifies that it can indeed have an overall subcritical effect, as well as giving its relative importance to other physical mechanisms.

We conclude this section by considering our results in relation to that of \citet{Kaplan2017}. In their study they find subcritical convection in a full sphere at low Prandtl numbers that comes with the creation of a strong zonal flow. They subsequently investigate the importance of this zonal flow for subcritical convection by artificially removing the zonal velocity and zonal temperature components from their simulation each time step. Even when the zonal flow is removed the subcritical solution is found to persist. It is interesting to compare our spherical shell results with these by considering the breakdown in table \ref{tab:coeff_int} which lets us isolate the effect of each term. We clearly see that a strong zonal flow is created due to the Reynolds stress term and thermal stress term (see the right hand side of equation (\ref{equ:qAAb})). If the zonal flow term is removed then the only term left that determines $\gamma$ is the harmonic term $\bm{q}_{AA}$. Our table shows that in this case subcritical convection would occur only for $\textit{Pr}=0.1$, $\textit{Ek}=10^{-7}$, and that the amount of {subcriticality}, as measured by $\gamma$, would be extremely weak. Hence, the axisymmetric flow induced through the Reynolds and thermal stresses is the most important term in determining whether convection onsets subcritically or not.

While this may seem in contradiction with the results of \citet{Kaplan2017}, it is simply a consequence of the differing areas of validity of our studies. Our current weakly nonlinear analysis gives an amplitude equation with a cubic nonlinearity. When $\textrm{Re}(\gamma)<0$, nonlinearities promote the growth of the instability and our amplitude equation will give unbounded growth. In order to determine how nonlinearities saturate this growth we would need to go to higher (fifth) order in our weakly nonlinear expansion, which is beyond the scope of this study. In other words, we cannot say anything about saturation in the subcritical case with our model, only that it is indeed subcritical. In contrast, the nonlinear study of \citet{Kaplan2017} naturally includes these higher order terms. Therefore, we should interpret their results as indicating that the zonal flow is not important for the saturation of the runaway subcritical growth. Consequently, once the convection has saturated, the zonal flow can be safely removed without inhibiting the convection. This is consistent with a weakly nonlinear analysis in which fifth order terms $|A|^4A$ would be dominant in this saturated regime, weakening the effect of the zonal flow. In light of our weakly nonlinear model we can predict that if the zonal flow was removed when the amplitude of the flow was small, meaning that the fifth order, and higher terms, are negligible in comparison to linear and cubic terms, that removing the zonal flow would suppress the convection. In this manner, our results are indicative of the weak branch of convection, rather than the strong branch of convection \citep{Kaplan2017}.

\section{Conclusions}\label{sec:conclusions}
We have presented the mathematical and numerical framework required for carrying out a weakly nonlinear analysis for the onset of convection in a spherical shell. By solving the weakly nonlinear equations, {and adjoint equations}, numerically we alleviate the difficulties associated with carrying out such an analysis analytically for a non self-adjoint system. We have carried out this procedure for an Earth-like radius ratio for a range of Ekman numbers and Prandtl numbers. The effects of two types of heating are considered; differential heating and internal heating. For our Boussinesq governing equations we show that once the onset of convection is determined, the weakly nonlinear amplitude equation coefficients can be determined by solving one 2D eigenvalue problem and two 2D linear boundary value problems, making this approach a very efficient method for determining the behaviour near {criticality} for an otherwise fully 3D nonlinear problem.

For differential heating we show that the weakly nonlinear Stuart--Landau amplitude accurately reproduces the results of a full 3D nonlinear simulation, verifying the framework. For all Ekman numbers considered the weakly nonlinear analysis reveals that convection onsets as a supercritical Hopf bifurcation. The dominant term, beyond the instability itself, in our weakly nonlinear expansion is a large zonal flow produced through Reynolds and thermal stresses. The velocity component of this zonal flow promotes {subcriticality}, whereas the stronger thermal component promotes {supercriticality} and leads to the overall supercritical behaviour of the instability. We show that the weakly nonlinear analysis can be used to predict the overall change in rotation rate of the instability due to nonlinearities as well as the amplitude and limit cycle behaviour obtained through saturation.

Conversely for internal heating subcritical convection is found for all Prandtl numbers considered provided the Ekman number is small enough. This subcritical convection is primarily obtained through the zonal temperature component changing to a subcritical effect for low enough Ekman numbers. For low Prandtl numbers the zonal velocity field becomes more dominant which can lead to subcritical behaviour even when the temperature component of the zonal flow is supercritical. The discovery of subcritical convection in a shell is in agreement with theoretical predictions \citep{soward_1977,plaut_lebranchu_simitev_busse_2008} and numerical studies for low Prandtl number convection in a full sphere \citep{Guervilly_2016,Kaplan2017}.

Overall our study has shown that weakly nonlinear analysis performed numerically is an extremely efficient method to determining whether an instability onsets subcritically or supercritically, as it requires the solution of 2D linear problems as opposed to 3D nonlinear calculations. Therefore, we propose going forwards that weakly nonlinear analysis can be used for determining the behaviour of the onset of convection in a variety of systems in parameter regimes inaccessible to direct numerical simulation. As the weakly nonlinear coefficients obtained can be broken down into their constituent terms, they can be used to probe the flow to reveal the dominant pathways governing the flow dynamics and determining the nature of the bifurcation. Further, we show that for a spherical shell the form of heating used is an important factor underlying whether convection onsets subcritically at low Ekman numbers. 

The weakly nonlinear analysis presented is easily extendable to other equations and geometries, and remains 2D as long as the instability contains at least one direction at a single wavenumber. A simple extension would be to consider different heating profiles and radius ratios. Different heating profiles could have more relevance to convection in other planetary bodies such as Mercury or Ganymede \citep{Jones_2011}, whereas different radius ratios would explore convection in the past and future Earth as well as other planetary bodies. Specifically, a high radius ratio, relevant to Earth's future, would be interesting as asymptotic theory indicates that convection in this case will behave significantly differently to a full sphere \citep{dormy_2004}. Extending our analysis to an anelastic system would allow for weakly nonlinear convection in highly stratified systems such as in gas giants to be determined. Furthermore, the more complicated nonlinearities present in anelastic equations provide more routes to saturation such as viscous heating, which would be uncovered through studying the weakly nonlinear terms. The reduced limit cycle description (\ref{equ:limitCycle}) could be used for studying secondary bifurcations, i.e. the stability of the weak convective branch to hydrodynamic perturbations could be studied without the use of Newton-Krylov methods to converge the weak branch limit cycles \citep{Garcia_2021}. As the reduced description clearly provides the zonal flow generated by the convection near {criticality}, it also provides a systematic way for finding and studying the zonal flow without conducting nonlinear 3D simulations. Finally, the effect of magnetic fields on these systems is a natural line of enquiry. Whilst magnetic fields cannot be self-consistently included in our weakly nonlinear expansion which is centered around a zero flow state that cannot produce a dynamo, the stability of magnetic fields on top of the reduced description of the limit cycle (\ref{equ:limitCycle}) obtained through weakly nonlinear analysis is possible in a manner similar to studying secondary hydrodynamic instabilities. {In this way, weakly nonlinear analysis can then be used to determine whether dynamos are supercritical or subcritical, and can also be used to determine the effects of imposed magnetic fields on the supercriticality or subcriticality of convection in spherical domains.}

\section*{Acknowledgements}
This work was undertaken on ARC4, part of the High Performance Computing facilities at the University of Leeds, UK. We would like to acknowledge Emmanuel Dormy and Andrew Soward for enlightening conversations. We acknowledge partial support from a grant from the Simons Foundation (Grant No. 662962, GF). We would also like to acknowledge support of funding from the European Union Horizon 2020 research and innovation programme (grant agreement no. D5S-DLV-786780).

\section*{Data availability statement}
The numerical code that supports the findings of this study is openly available \citep{SkeneTobias_code2024}.

\section*{Disclosure statement}
The authors report there are no competing interests to declare.

\bibliographystyle{unsrtnat}


\end{document}